\documentclass[a4paper,11pt]{article}
\pdfoutput=1

\usepackage{jheppub}
\usepackage{amsmath,amssymb,amsfonts}
\usepackage{graphicx}
\usepackage{cleveref}
\usepackage{hyperref}
\usepackage{srcltx}
\usepackage{bm}
\usepackage{mathdots}
\usepackage{color}
\usepackage{dsfont}
\usepackage{slashed}
\usepackage{physics}
\usepackage{subfigure}
\usepackage{tikz}
\usepackage[normalem]{ulem}
\usepackage{orcidlink}
\usepackage{enumitem}
\usepackage{url}
\usepackage{comment}
\usepackage{cancel}
\usepackage{commath}
\usepackage{tikz}
\usepackage[compat=1.1.0]{tikz-feynman}
\usepackage{placeins}

\title{Near-SUSY to Non-SUSY Crossover}

\author[a]{Dan Kondo\,\orcidlink{0000-0002-6268-3332},}\emailAdd{dan.kondo@ipmu.jp}
\author[a,b,c,1]{Hitoshi Murayama\,\orcidlink{0000-0001-5769-9471}\note{Hamamatsu Professor}}\emailAdd{hitoshi@berkeley.edu}

\author[b]{and Bea Noether\,\orcidlink{0000-0002-2947-3210}}	\emailAdd{bea\_noether@berkeley.edu}

\affiliation[a]{Kavli Institute for the Physics and Mathematics of the Universe (WPI), University of Tokyo Institutes for Advanced Study, University of Tokyo, Kashiwa 277-8583, Japan}
\affiliation[b]{Leinweber Institute for Theoretical Physics, University of California, Berkeley, CA 94720, USA}
\affiliation[c]{Ernest Orlando Lawrence Berkeley National Laboratory, Berkeley, CA 94720, USA}

\abstract{
Gauge theories can be solved exactly slightly away from the supersymmetric (SUSY) limit softly broken by anomaly mediation when the size of SUSY breaking is much smaller than the dynamical scale ($m \ll \Lambda$). We show empirical evidence that the near-SUSY limit is continuously connected to the non-SUSY limit ($m \gg \Lambda$) in $\mathrm{SU}(N_c)$ gauge theories with $N_f$ quarks in the fundamental representation. The evidence includes the behavior of quark bi-linear condensate and gluon condensates, light hadron spectra, and consistency with the large $N_c$ limit including the Witten-Veneziano relations between the topological susceptibility and $m_{\eta'}$ and $m_\pi$. In addition, we present new predictions when $N_f/N_c \gtrsim O(1)$. }

\begin{document}
\maketitle
\flushbottom
\setcounter{page}{2}

\section{Introduction}

Understanding strong interaction in particle physics has been a challenging process. When Chadwick discovered the neutron~\cite{Chadwick:1932wcf}, it quickly became clear that all atoms were made of nuclei consisting of protons and neutrons surrounded by a cloud of electrons. Yet it was a major mystery how nuclei could bind despite the Coulomb repulsion among protons. Yukawa hypothesized pions to hold protons and neutrons together~\cite{Yukawa:1935xg}, which was later discovered in cosmic-ray experiments with much confusion with muons of a similar mass~\cite{Neddermeyer:1937md}. It was puzzling, however, why pions are so much lighter than nucleons so that the binding force extends beyond their size for the nucleons to be bound. Nambu and Jona-Lasinio argued that the pions are light because of the spontaneously broken chiral symmetry~\cite{Nambu:1961tp,Nambu:1961fr}, drawing from an analogy to the superconductors that spontaneously break the $\mathrm{U}(1)$ gauge symmetry of electric charge by the condensation of Cooper pairs in the Bardeen--Cooper--Schrieffer (BCS) state~\cite{Bardeen:1957kj}.

But it was only the beginning of the discovery of zoo of strong-interacting particles collectively called hadrons. We can now count more than one hundred hadrons listed by Particle Data Group~\cite{ParticleDataGroup:2024cfk}. The spectrum of light hadrons appeared to have a linear relationship between their spin and mass-squared in Chew-Fraustschi plots~\cite{Chew:1962eu}, which may be explained with a linear potential between constituents. Gell-Mann proposed an organizational principle based on the $\mathrm{SU}(3)$ flavor symmetry~\cite{Gell-Mann:1961omu}, which eventually led to the hypothesis of quarks $u, d, s$ with fractional charges as their fundamental constituents by Gell-Mann and Zweig~\cite{Gell-Mann:1964ewy,Zweig:1964ruk}. Nambu proposed that the fundamental constituents have color degrees of freedom bound together by color-octet gluons~\cite{Han:1965pf}. Eventually they were put together in Quantum ChromoDynamics (QCD) with non-abelian gauge theory based on the $\mathrm{SU}(3)$ color group with three flavors $u, d, s$ by Fritzsch, Gell-Mann, and Leutwyler~\cite{Fritzsch:1973pi}. 

On the other hand in 1968, deep inelastic scattering (DIS) experiments at SLAC revealed nearly free ``partons'' inside proton, the name coined by Feynman~\cite{Feynman:1969ej}. Gross and Wilczek \cite{Gross:1973id}, and independently Politzer \cite{Politzer:1973fx}, discovered that non-abelian gauge theories can exhibit asymptotic freedom, allowing the fundamental constituents to behave like free particles at higher energies. Yet the idea of fractionally charged constituents and the confinement that does not allow quarks to be liberated were the major stumbling block for the acceptance of quarks and gluons. Only after $J$ and $\psi$ particles were independently discovered at Brookhaven \cite{E598:1974sol} and SLAC~\cite{SLAC-SP-017:1974ind,Abrams:1974yy}, respectively, followed by discovery of charmed mesons~\cite{PhysRevLett.37.255}, quarks bound by gluons became the accepted picture of all hadrons. 

Yet the idea of confinement by a linear potential that the fundamental constituents can never be liberated and observed in isolation, as well as that the dynamics spontaneously breaks the chiral symmetry, proved elusive for theoretical understanding. Mandelstam proposed that the confinement might be understood as the ``dual Meissner effect.'' He noted that the magnetic flux is squeezed in the Abrikosov flux tube in Type-II superconductor due to the condensate of Cooper pairs~\cite{Mandelstam:1974pi}. The energy of magnetic flux tubes is proportional to the length, and it would give rise to a linear potential between a magnetic monopole and an anti-magenetic monopole if they are embedded inside the superconductor. Considering the electric-magnetic duality, he argued that there should be a linear potential among charged particles if there is a condensate of magnetic monopoles. But it was not clear if there are magnetic monopoles in the $\mathrm{SU}(3)_C$ gauge theory, and how they are connected to the dynamics of chiral symmetry breaking.

Much later, formulation of QCD as a path integral on a latticized Euclidean spacetime allowed for a breakthrough~\cite{Wilson:1974sk}. We can now show the linear potential, or equivalently the area law of the Wilson loop, in QCD without dynamical quarks~\cite{degrand2006lattice,Gattringer:2010zz,lellouch2011modern}. In the case with dynanical quarks, which is still a major technical challenge in lattice QCD, simulations show that there are bilinear condensates of quarks as order parameters of chiral symmetry breaking. However the numerical computer simulations do not provide us better insight into the basic understanding of both confinement and chiral symmetry breaking. In addition, it remains a technical challenge to define chiral symmetry, include a large number of dynamical fermions, deal with larger gauge groups. Chiral gauge theories such as standard model  cannot be simulated still today. 

In 1990s, Seiberg and Witten showed how $N=2$ supersymmetric gauge theories can be solved exactly, and demonstrated confinement by monopole condensation upon perturbation to $N=1$~\cite{Seiberg:1994rs}. Yet $N=2$ theories do not accommodate full chiral symmetry we need to study QCD. Subsequently, Seiberg, Intriligator, and Pouliot worked out $N=1$ vector-like gauge theories~\cite{Seiberg:1994bz,Intriligator:1994uk,Intriligator:1994jr,Intriligator:1995ne,Intriligator:1995au}. They exhibited a wide range of phases from no ground state with run-away behavior, infinite number of vacua with quantum constraints, confinement without chiral symmetry breaking, and exotic magnetic theories and superconformal dynamics. Unfortunately, none of them are similar to what we expect in non-SUSY QCD.

More recently, beginning from~\cite{Murayama:2021xfj}, the applications of soft-supersymmetry breaking effects via anomaly mediation to the solutions to $N=1$ vector-like and chiral gauge theories~\cite{Csaki:2021xhi,Csaki:2021aqv,Csaki:2021jax,Csaki:2021xuc,Kondo:2022lvu,Leedom:2025mcg,Goh:2025oes}, allows for exact solutions as long as the supersymmetry breaking effects are much smaller than the dynamical scale $m \ll \Lambda$. They resembled very much what we expect in the non-SUSY limits $m \rightarrow \infty$. While a rigorous proof of the absence of a phase transition remains challenging \cite{Dine:2022req,Dine:2022nmt}, it suggested that changing $m$ from near-SUSY limit $m \ll \Lambda$ to non-SUSY $m \rightarrow \infty$ is continuous without a phase boundary and hence is a crossover.

In this paper, we present more evidence that the near-SUSY and non-SUSY limits belong to the same universality class and it is a crossover. This behavior is akin to the BCS--BEC cross over in cold atoms. In particular, we show evidence in three different areas:
\begin{enumerate}
    \item Correct scaling of $f_\pi^2 \sim O(N_c)$ in the large $N_c$ limit. This point was not obvious since a paper \cite{Martin:1998yr} concluded that $f_\pi^2 \sim O(N_c)^0$ in this case and argued that there must be a phase transition where $m \sim \Lambda$. We demonstrate that this is not the case. We show the behavior of $f_\pi^2$ when $N_f / N_c \gtrsim O(1)$ for the first time. We also reproduce the relationship between the topological susceptibility and the masses $m_\pi^2$ and $m_{\eta'}^2$ at large $N_c$ explicitly.
    
    \item Consistency of non-perturbative condensates. We show how even quark bilinear $\langle \bar{q}q\rangle $ and gluon $\langle G_{\mu\nu} G^{\mu\nu}\rangle$ condensates can be computed and that they satisfy a non-trivial consistency between the $m \ll \Lambda$ and $m \gg \Lambda$ regimes, and could plausibly be continuously connected. Moreover they are consistent with expected large $N_c$ behavior when $N_f \ll N_c$ while we also work out analytically their behavior when $N_f / N_c \gtrsim O(1)$ for the first time.
    \item Low-lying spectrum. Not only that the massless pions persist, we even computed the case with light quarks, and not only pseudoscalars $J^P=0^-$ but also scalars $0^+$ appear qualitatively (and even semi-quantitatively) consistent between the near-SUSY and non-SUSY limits.
\end{enumerate}

The organization of the paper is as follows. In~\cref{sec:BCSBECcrossover}, we briefly review the notion of BCS-BEC crossover to help for understanding the notion of the crossover. In~\cref{sec:vacua}, we review on the vacuum structure of SQCD deformed by AMSB (ASQCD) similar to~\cite{Csaki:2022cyg} with updated results. In~\cref{sec:chiralLagrangianADS}, we derive the chiral Lagrangian and Wess-Zumino-Witten term in the ADS case. In~\cref{sec:condensate}, we calculate the various condensates. In~\cref{sec:massspectrum}, we show the mass spectrum for ADS, quantum modified, s-confinement phase.

\section{BCS--BEC Crossover}\label{sec:BCSBECcrossover}

In this section, we briefly review the concept of BCS-BEC crossover (see {\it e.g.}\/, \cite{parish2015bcs,chen2024superconductivity} for reviews), and its analogy to the supersymmetric QCD. 

The BCS--BEC crossover concerns a system of fermions at the zero temperature. When there is no attractive force among fermions $g=0$, it is a degenerate Fermi gas with unbroken $\mathrm{U}(1)$ symmetry for the conservation of number of particles. On the other hand, once $g$ is turned on, the Fermi surface becomes unstable and the ground state becomes the BCS state \cite{Bardeen:1957mv} where the fermion bilinear has a condensate which spontaneously breaks the $\mathrm{U}(1)$ symmetry. We can work out quantitative details using the variational ground state with perturbation in small couplings. As the coupling $g$ is increased, analytical study becomes impossible. Yet numerical simulations show that the $\mathrm{U}(1)$ symmetry remains broken. Once $g$ is very large, we arrive at a different description of the system. Namely that fermions form a bosonic bound state, and this bound state boson undergoes Bose--Einstein Condensation (BEC). There is no phase transition as a function of $g$ and hence this phenomenon is called BCS--BEC crossover. Note that the point $g=0$ is special with no symmetry breaking, while other points with $g>0$ share the same universality class.

\begin{figure}
    \centering
    \includegraphics[width=0.75\linewidth]{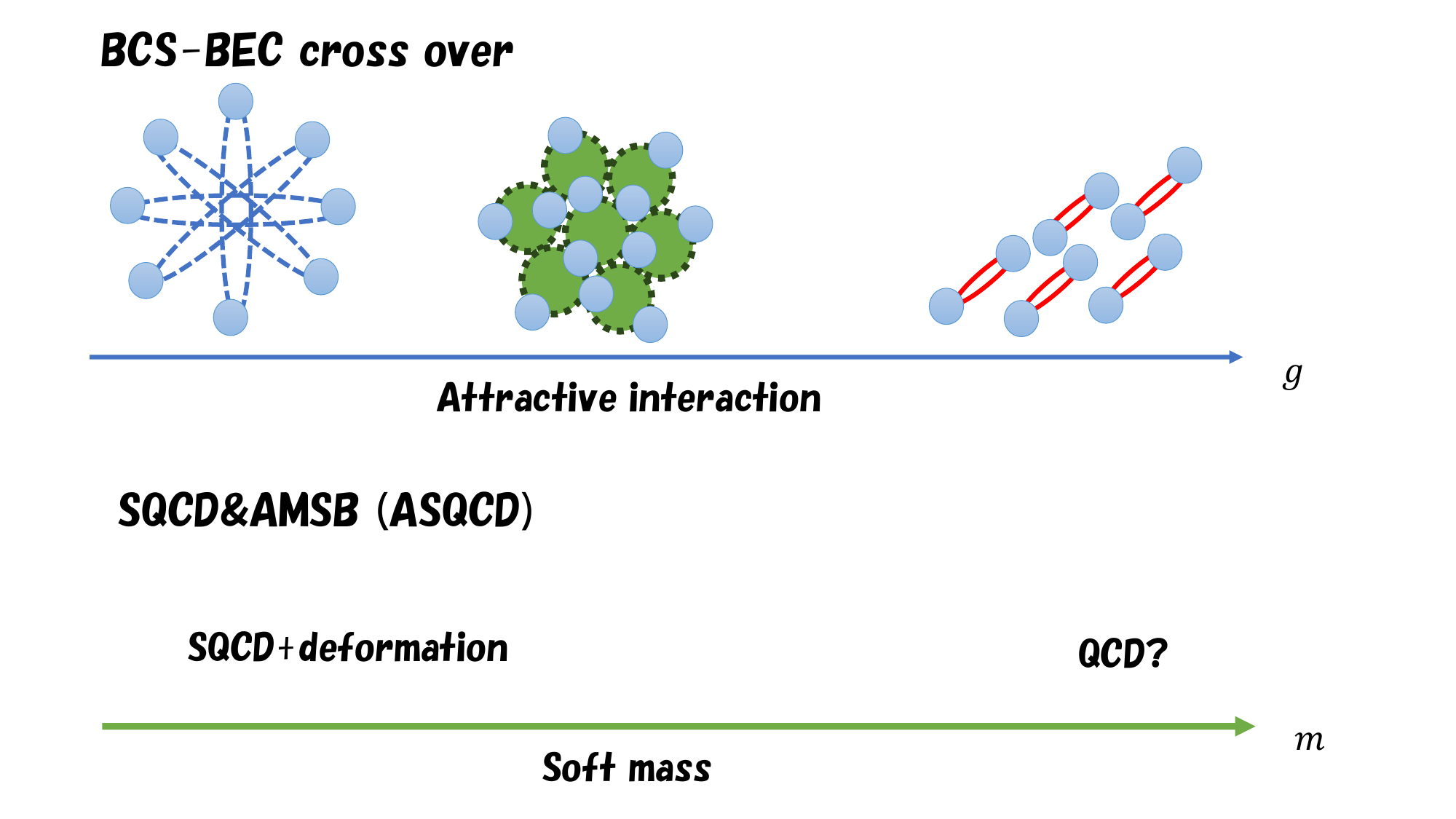}
    \caption{The illustration of BCS--BEC crossover, by changing the attractive interaction $g$. Only the left-most point $g=0$ does not break the $\mathrm{U}(1)$ symmetry, while from the BCS state with small $g$ to the BEC with infinite $g$ is a crossover and the universality class of broken $\mathrm{U}(1)$ remains unchanged. We expect a similar crossover for the deformation of near-SUSY SQCD ($m\ll \Lambda$) into non-SUSY QCD ($m\rightarrow \infty$), while SQCD itself ($m=0$) belongs to a different universality class. }
    \label{fig:BCSBEC}
\end{figure}

We show a schematic picture of the BCS-BEC crossover in the upper part of \cref{fig:BCSBEC}. 


In this paper, we will argue and show some evidence that a similar crossover occurs in supersymmetric $\mathrm{SU}(N_c)$ QCD with $N_f$ vector-like quarks in the fundamental representation, as illustrated in the lower part of \cref{fig:BCSBEC}. The SUSY limit $m=0$ is the famous exact result obtained by Seiberg \cite{Seiberg:1994bz}, whose ground states do not resemble the non-SUSY limit at all which is expected to be either chiral symmetry breaking $\mathrm{SU}(N_f)_L \times \mathrm{SU}(N_f)_R \rightarrow \mathrm{SU}(N_f)_V$ or an IR fixed point. 


For a relatively small number of quark flavors, when a small soft SUSY breaking is turned on via anomaly mediation $m \ll \Lambda$, we can still solve the theories exactly and show that the chiral symmetry breaking occurs. In this near-SUSY limit, we can {\it derive}\/ the chiral Lagrangian, its Wess--Zumino--Witten term, and various non-perturbative condensates including quark bilinear and gluon condensates quantitatively. This is our ``BCS'' theory. As $m$ is increased to $m \sim \Lambda$, we lose our theoretical control. Once $m \gtrsim \Lambda$, we need to switch our description based on composite pions with its chiral Lagrangian. This is our ``BEC'' theory. Yet the non-SUSY limit $m \gg \Lambda$ with the standard chiral symmetry breaking appears to be continuously connected without a phase transition, so that the theories remain in the same universality class. This is true for all $N_f < 1.48 N_c$.

For larger $N_f$, what is continuously connected may be a local minimum in the near-SUSY limit, while the global minimum breaks $\mathrm{U}(1)_B$. They switch when $m \simeq \Lambda$. Namely there is phase transition. But it is possible to follow the local minimum from $m \ll \Lambda$ which appears to be continuously connected to the non-SUSY limit. Therefore, studying the near-SUSY limit on a local minimum is still useful to gain insight into the non-perturbative dynamics. In particular, we can work out the non-perturbative condensates in the large $N_c$ limit with comparably large $N_f$ which has not been possible with other methods. 

\section{Chiral Symmetry Broken vacuum in ASQCD}\label{sec:vacua}
In this section, we comprehensively review the vacuum configuration of ASQCD, which we will use the results later. For the calculational details, you can refer to~\cite{Murayama:2021xfj,Csaki:2022cyg,Kondo:2021osz}.

  \subsection{Affleck-Dine-Seiberg (ADS) case}
In the ADS case where $N_f<N_c$, the dynamics is described by meson fields $M_{ij}$ with superpotential 
\begin{align}\label{eq:WADS}
    W&= (N_c-N_f)\left(\frac{\Lambda^{3N_c-N_f}}{\det M}\right)^{\frac{1}{N_c-N_f}}
\end{align}
Meson is in the form of $M_{ij}=M\delta_{ij}$, which is D-flat direction.
\begin{align}
    Q=\tilde{Q}=
    \begin{pmatrix}
        1&0&\cdots&0\\
        0&1&\cdots&0\\
        \vdots&\vdots&\ddots&\vdots\\
        0&0&\cdots&1\\
        0&0&\cdots&0\\
        \vdots&\vdots&\ddots&\vdots\\
        0&0&0&0
    \end{pmatrix}\phi,\ 
    M=\phi^2.
    \label{eq:ADSvevs}
\end{align}
Whereas the pure SUSY theory has no minimum, the addition of AMSB creates a stable minimum at
\begin{align}
    \phi^2 =& \Lambda^2 \left(
    \frac{
    N_c+N_f}{3N_c-N_f} \frac{\Lambda}{m}
    \right)^{\frac{N_c-N_f}{N_c}}.\label{eq:ADSVEV}
\end{align}

\subsection{Quantum modified moduli space case}
Quantum modified phase is where $N_f=N_c$. We are interested in the region where $\chi$SB vacuum is realized, where the meson $M$ has $\chi$SB vacuum expectation value, while the baryon $B$ has vanishing value. In canonical and minimal K\"ahler potential, we have the superpotential with a Lagrange multiplier $X$,
\begin{align}\label{eq:WQM}
    W&= \lambda X\left(\frac{\det M}{\Lambda^{N_c-2}}-\Lambda^2\right).
\end{align}
We find the minimum, at next-to-leading order in $m$, is at 
\begin{align}
    M_{ij}&= \Lambda\left(1 - \frac{m^2}{N_c \lambda^2 \Lambda^2} + O(m^4) \right) \delta_{ij}, \label{eq:QMVEV}\\
    X&= \frac{m}{\lambda}+\frac{m^3}{\lambda^3\Lambda^2}
    + O(m^5).
\end{align}
We note that the minimum here seems different from the literature like in \cite{Murayama:2021xfj}. This is because we absorbed factors of $\lambda$ into the constant $\Lambda$ and use a different form of the superpotential than that used in the literature. The final result with $\chi$SB vacuum is consistent.

Unfortunately for this case, the expectation values are at the strong scale and the K\"ahler potential for the meson and baryon fields may have arbitrary higher order terms in fields.\footnote{However, it is not possible that there is ``baryonic run-away" \cite{Luzio:2022ccn} because the potential is stablized by the two-loop AMSB effects $m_{Q}^2 = m_{\tilde{Q}}^2 > 0$ once $Q,\tilde{Q} \gg \Lambda$.}  In fact, it is not possible to exclude the possibility that the baryon fields also acquire expectation values. We will simply assume that the above vacuum is the true vacuum for the sake of discussions below.

\subsection{s-confinement case}
In the $s$-confining phase $(N_f=N_c+1)$, the superpotential is
	\begin{align}
		W =&(N_f-N_c)\left(\frac{\det M}{\Lambda^{3N_c-N_f}}\right)^{\frac{1}{N_f-N_c}}+\frac{1}{\mu} \tilde{B} M B.
	\end{align}
We wrote it in a suggestive fashion that the meson part extends to the free magnetic phase. We rescale the scalar field to the canonical normalization $M\rightarrow\Lambda M$,
\begin{align}\label{eq:Wsconfine}
		W =&(N_f-N_c)\left(\frac{\det M}{\Lambda^{3N_c-2N_f}}\right)^{\frac{1}{N_f-N_c}}+\frac{\Lambda}{\mu} \tilde{B} M B.
	\end{align}
We find that the vacuum conserves the baryon number $B=\tilde{B}=0$. The potential is
\begin{align}
    V&=\sum_{i=1}^{N_f}\left|\frac{1}{M_i}\frac{\det M}{\Lambda^{3N_c-2N_f}}\right|^2-2m(N_f-3)\frac{\det M}{\Lambda^{3N_c-2N_f}}\ .
\end{align}
In the chiral symmetry broken vacuum $M=\text{diag}(\phi,\cdots,\phi)$ and the scalar potential will be
\begin{align}
    V =& N_f \frac{\phi^{2N_f-2}}{\Lambda^{2N_f-6}}-2(N_f-3)m\frac{\phi^{N_f}}{\Lambda^{N_f-3}}\ .
\end{align}
We have the minimum in the diagonal meson direction $M^{ij} = \phi\delta^{ij}$ and $B=\tilde{B}=0$ with
	\begin{align}
		\phi =& \Lambda\left(\frac{N_f-3}{N_f-1}\frac{m}{\Lambda}\right)^{\frac{1}{N_f-2}}.\label{eq:s-confineVEV}
	\end{align}

\subsection{Free Magnetic case}
Free magnetic phase is the region where $N_c+1<N_f\leq \frac{3}{2}N_c$. The effective superpotential after integrating out the magnetic quarks in the presence of ${\rm det}M \neq 0$ is
	\begin{align}\label{eq:WFM}
		W =& (N_f-N_c)\left(\frac{\kappa^{N_f}\phi^{N_f}}{\Lambda^{3N_c-2N_f}}\right)^{\frac{1}{N_f-N_c}},
	\end{align}
where the scalar potential in the direction $M_{ij}=\kappa \Lambda \phi\delta_{ij}$ is
\begin{align}
    V =& N_f \Lambda^4 \abs{\frac{\kappa\phi}{\Lambda}}^{\frac{2N_c}{N_f-N_c}} - (3N_c-2N_f)m\Lambda^3 \left(\frac{\kappa\phi}{\Lambda}\right)^{\frac{N_f}{N_f-N_c}} + c.c.
\end{align}
We obtain the minimum at
	\begin{align}
		\phi =& \frac{\Lambda}{\kappa}\left(\frac{3N_c-2N_f}{N_c}\frac{m}{\Lambda}\right)^{\frac{N_f-N_c}{2N_c-N_f}}.\label{eq:FreeMagVEV}
	\end{align}

Note, however, that the $\chi$SB is absent when $N_f = \frac{3}{2} N_c$. This is because the suprepotential \cref{eq:WFM} is classically conformal and AMSB is not effective. It is considered as a limitation of the method which cannot be overcome even including two loop and three loop contributions \cite{deLima:2023ebw}. Here we take a pragmatic attitude that this particular case is an ``anomaly'' in this method beyond our scope, and focus on how $N_f < \frac{3}{2}N_c$ and $N_f > \frac{3}{2}N_c$ shows a continuity in the large $N_c$ limit where $N_f$ can be considered as a continuous variable.

\subsection{Conformal Window}
In SQCD, the conformal window is where $\frac{3}{2}N_c < N_f < 3N_c$. The SUSY theory in this range flows to SCFT in the IR. Both the ``Electric" $\mathrm{SU}(N_c)$ theory and its Seiberg dual, a ``Magnetic" $\mathrm{SU}(N_f-N_c)$ theory with an additional singlet meson, describe the same physics in the IR limit. In \cite{Kondo:2021osz} it was shown that the AMSB is a relevant perturbation and deflects the theory from the IR fixed point. If one assumes Seiberg duality is valid sufficiently close to the fixed-point, these conclusions apply to ASQCD in the (now ``broken") conformal window range of $N_f$. In particular, magnetic theory is more appropriate near $N_f=\frac{3N_c}{2}$ and electric theory is more appropriate near $N_f=3N_c$. \\

The Magnetic theory has the following non-perturbative superpotential in the IR, using $\tilde{N}_c\equiv N_f-N_c$, after integrating out the dual quarks 
	\begin{align}
		W =& \tilde{N}_c\left(\tilde{\Lambda}^{3\tilde{N}_c-N_f}\det\frac{M}{\mu_m}\right)^{\frac{1}{\tilde{N}_c}}.
	\end{align}
Once the theory reaches the IR fixed point, this effective superpotential is conformal and the AMSB is not effective. However, at least near the lower edge of the conformal window, it was shown that AMSB effects are relevant and the theory does not quite reach the IR fixed point \cite{Kondo:2021osz}.

The QCD-like minimum of the AMSB-perturbed magnetic theory, near the lower edge of the window ($N_f\simeq \frac{3N_c}{2}$), is given by $M_{ij}=\phi\delta_{ij}$ with
\begin{align}
\phi =& \tilde{\Lambda}\left(\frac{\alpha(2+\alpha) }{16 \lambda^{\frac{2N_f}{\tilde{N}_c}} N_f^{\frac{2\tilde{N}_c-N_f}{\tilde{N}_c}} }\right)^{\frac{1}{2-\alpha}}\left(\frac{m}{\tilde{\Lambda}}\right)^{\frac{2}{2-\alpha}}, \label{eq:LCWVEV}\\
\tilde{\epsilon}=&\frac{3\tilde{N}_c-N_f}{N_f},\ 
\alpha=21\tilde{\epsilon}^2,
\end{align}
where unknown $\mathcal{O}(1)$ constants have been set to 1, $\tilde{\Lambda}$ is the holomorphic scale of the Magnetic theory, and $\lambda$ is the Yukawa coupling. 

The Twice-Dual theory features the quark superfields $Q_i,\tilde{Q}_i$, as well as the meson $M_{ij}$ and an additional singlet ``dual-dual" meson $N_{ij}$. This is a way we can capture the full rank of the meson matrix and the deviation away from the IR fixed point in this case proposed in \cite{Kondo:2021osz}. The superpotential is
\begin{align}
    W =& Y N(Q\tilde{Q}-M),
\end{align}
where $Y$ is a Yukawa coupling. The additional meson $N$ serves as a Lagrange multiplier in the SUSY theory, enforcing $M=Q\tilde{Q}$ and $W=0$, reproducing the original electric theory. In the AMSB theory, however, the scalar component of $N$ can acquire a VEV and we can (sufficiently close to the upper edge of the window) calculate the minimum of the theory. This is in contrast to the original electric theory, which perturbed by AMSB becomes strongly coupled and incalculable in the IR. 

The QCD-like minimum of the AMSB-perturbed twice-dual theory, near the upper edge of the window, is given by $M_{ij}=\phi\delta_{ij}$ where
\begin{align}
\phi =& \Lambda\left(\frac{m}{Y \Lambda}\right)^{\frac{1}{\epsilon}},\label{eq:UCWVEV}
\\
\epsilon =& \frac{3N_c-N_f}{N_f}\ ,
\end{align}
where again unknown $\mathcal{O}(1)$ coefficients have been set to 1. 

For intermediate values $\frac{3}{2}\ll \frac{N_f}{N_c} \ll 3$ the couplings of either description near the fixed point become large, and we lose the ability to calculate the minimum of the AMSB-perturbed theory. We can only interpolate the results from the edges into this region based on the assumption of continuity.

\section{Derivation of Chiral Lagrangian and Wess-Zumino-Witten term}\label{sec:chiralLagrangianADS}

In this section, we {\it derive}\/ the chiral Lagrangian as the low-energy limit of SQCD with a small AMSB. 

\subsection{Chiral Lagrangian in the ADS phase}\label{subsec:chiralLagADS}
We derive chiral Lagrangian from the first principle. A similar analysis was done in \cite{Csaki:2023yas} but without the hidden local symmetry.

We will parametrize the coset space 
\begin{align}
    Q&= \begin{pmatrix}
        v\xi^T\\0
    \end{pmatrix},\ 
    \tilde{Q}= \begin{pmatrix}
        v\xi\\0
    \end{pmatrix},
    \label{eq:parametrization}
\end{align}
where $\xi$ transforms $\xi\rightarrow g_L\xi h^T$ under $\mathrm{SU}(N_f)_L$ and $\xi\rightarrow h^*\xi g_R^T$ under $\mathrm{SU}(N_f)_R$. Here, $h$ is the $\mathrm{SU}(N_c)$ gauge transformation needed to keep the $Q$ and $\tilde{Q}$ fields in the form \cref{eq:parametrization} after the global transformations $(g_L,g_R)$. Note that $U\equiv \xi\xi=\xi^2$ transforms $U\rightarrow g_LUg_R^T$ under a simultaneous transformation of $\mathrm{SU}(N_f)_L\times \mathrm{SU}(N_f)_R$. Then from kinetic term, we can obtain
\begin{align}\label{eq:Qkin}
    \mathcal{L}_{\text{kin}}
    &=\operatorname{Tr}|D_\mu Q|^2+\operatorname{Tr}|D_\mu \tilde{Q}|^2\nonumber\\
    &=-\frac{v^2}{2}\operatorname{Tr}(\xi\partial_\mu\xi^\dag-\xi^\dag\partial_\mu\xi)^2+\frac{v^2}{2}\operatorname{Tr}(2g\rho_\mu-i(\xi^\dag\partial_\mu\xi+\xi\partial_\mu\xi^\dag))^2\nonumber\\
    &=\frac{v^2}{2}\operatorname{Tr} \partial_\mu U^\dagger \partial^\mu U+\frac{v^2}{2}\operatorname{Tr}\left(2g\rho_\mu-\frac{i}{2}(U^\dag \partial_\mu U+U\partial_\mu U^\dag)\right)^2.
\end{align}
$\rho_\mu$ is the $\mathrm{SU}(N_c)$ gauge fields in the $N_f \times N_f$ block. It is clear that this form of the low-energy Lagrangian is consistent with the ``hidden local symmetry'' \cite{Bando:1984ej,Bando:1985rf}. See Appendix~\ref{app:kinetic} for more details.

We can also start from the K\"ahler potential written with the meson field. The corresponding parameterization of the meson is,
\begin{align}
    M=\frac{v}{\sqrt{2}}U,
\end{align}
which transforms in the same way as above $U\rightarrow g_LUg_R^T$ under $\mathrm{SU}(N_f)_L \times \mathrm{SU}(N_f)_R$. The meson kinetic term is
\begin{align}
    \operatorname{Tr}\partial_\mu M^\dagger \partial^\mu M=\frac{v^2}{2}\operatorname{Tr} \partial_\mu U^\dagger \partial^\mu U.
\end{align}
In either way, we obtain the same kinetic term and decay constant.

We can read out the decay constant $f_\pi$ and mass of the $\rho$ meson from~\cref{eq:Qkin},
\begin{align}
    f_\pi^2&=2v^2 ,\\
    m_{\rho}^2&=2g^2v^2=g^2f_\pi^2.
\end{align}
Compared to \cite{Bando:1984ej,Bando:1985rf}, with parameterization
\begin{align}
    \mathcal{L}&= \mathcal{L}_A+a\mathcal{L}_V-\frac{1}{4g^2}F^{(\rho)}_{\mu\nu}F^{(\rho)\mu\nu}, \\
    \mathcal{L}_A&= \frac{f_\pi^2}{4}\operatorname{Tr} \partial_\mu U^\dagger \partial^\mu U, \\
    \mathcal{L}_V&= f_\pi^2 \operatorname{Tr}(V_\mu-\frac{i}{2}(\xi^\dag\partial_\mu\xi+\xi\partial_\mu\xi^\dag))^2,
\end{align}
under assumption that $\rho$ kinetic term is dynamically generated, our case corresponds to the choice $a=1$. Note that
the kinetic term of $\rho$ is derived in our case, not an assumption. In their case, they put a parameter $a$ because the vector term $\mathcal{L}_V$ is an auxiliary, which vanishes by equation of motion. It is often said that the choice $a=2$ reproduces the KSRFI\hspace{-1.2pt}I relation $m_\rho^2=2g_{\rho\pi\pi}^2f_\pi^2$ with $g_{\rho\pi\pi}=ag/2$. But arbitrary choice of $a$ reproduces the KSRFI relation $g_\rho/g_{\rho\pi\pi}=2f_\pi^2$ \cite{Bando:1985rf}, and our case ($a=1$) also holds it. 

It is not clear if $a=2$ is actually required for phenomenology. In the study of non-Linear Sigma model with large $N$ in \cite{Yamawaki:2018jvy,Yamawaki:2023ybn},  KSRFI\hspace{-1.2pt}I relation and Vector Meson Dominance (VMD) also hold for arbitrary $a$. They calculated the infinite sum of bubble diagram and compare the mass of $\rho$ meson to check the relation between the mass and the coupling. Also, they calculated the scattering amplitude of $\rho\pi\pi$, the $a$ dependent term is in the contact term, the tree and one loop contribution from the contact term (hence $a$ dependence) cancel each other and show the VMD. We quote their statement here: ``{\it the choice of $a=2$ is just an artifact of tree level effect}\/." In that sense, our case $a=1$ is a natural consequence that can capture the phenomenology of $\rho$ meson (at least for large $N$). They are $\rho$ meson coupling universality, KSRFI, KSRFI\hspace{-1.2pt}I relation and VMD. We regard this issue as an open question for the strongly coupled limit.

\subsection{Chiral Lagrangian for $N_f > N_c$}

For $N_f > N_c$, the description with the fundamental quark fields $Q$ and $\tilde{Q}$ is no longer valid. On the other hand, we have shown that the meson field acquires an expectation value. The K\"ahler potential for the meson field is what gives the chiral Lagrangian. Namely that with the minimal K\"ahler potential,
\begin{align}
    \int d^4 \theta {\rm Tr} M^\dagger M
    = {\rm Tr} \partial_\mu M^\dagger \partial^\mu M.
\end{align}
We introduce the chiral field as
\begin{align}
    M = \frac{f_\pi}{2} U.
\end{align}
Then we can read off 
\begin{align}
    f_\pi^2 = 4 \frac{1}{N_f} {\rm Tr} \langle M \rangle^\dagger \langle M \rangle.
\end{align}

In the s-confinement and free magnetic phase, the meson is weakly interacting and this is a justified derivation of the chiral Lagrangian. In the conformal window, this derivation is valid at the lower and upper edge of the window, which extends to the entire conformal window assuming the continuity in $N_f/N_c$. We regard this derivation highly plausible.

\subsection{Color Factor: 1PI coupling vs holomorphic coupling}
It is important to consider the difference of the coupling between holomorphic coupling $g_h$ and 1PI coupling $g_c$. They are related by the rescaling anomaly \cite{Arkani-Hamed:1997lye}
\begin{align}
    \frac{8\pi^2}{g_h^2}&= \frac{8\pi^2}{g_c^2}+C_A\log g_c^2,
\end{align}
where $C_A = N_c$ is adjoint Casimir. (Here we are ignoring another correction that stems from the Konishi anomaly due to the wave function renormalization factor $Z$ which is not important here.)

The authors of the paper \cite{Martin:1998yr} claimed that softly-broken SQCD and QCD are disconnected with a phase transition from the non-SUSY limit, as evidenced by an incorrect scaling of $f_\pi^2 \sim N_c^0$ they found when $N_f < N_c$. However, it seems that the authors did not relate the dynamical scale to the 't Hooft coupling correctly. We find explicit results for $f_\pi^2$ in ASQCD for all values of $N_f<3N_c$. In the particular case of $N_f<N_c$ and $N_c\gg 1$, we find that the correct $f_\pi^2\propto N_c$ scaling is reproduced. We also show new results for the scaling of $f_\pi^2$ in regimes where $N_f$ is not small compared to $N_c$.

For SQCD,
\begin{align}
    \Lambda^{3N_c-N_f}
    &=\mu^{3N_c-N_f}\mathrm{e}^{-8\pi^2/g_h^2}
=\mu^{3N_c-N_f}(g_c^2)^{-N_c}\mathrm{e}^{-8\pi^2/g_c^2}.
\end{align}
With 't Hooft coupling $N_cg_c^2=g_t^2$, this becomes
\begin{align}
    \Lambda^{3N_c-N_f}
&=\mu^{3N_c-N_f}(g_c^2)^{-N_c}\mathrm{e}^{-8\pi^2/g_c^2}
=\mu^{3N_c-N_f}N_c^{N_c}(g_t^2)^{-N_c} \mathrm{e}^{-8\pi^2N_c/g_t^2}.
\end{align}
We found that for $N_f<N_c$
\begin{align}
    f_\pi^2& = 2\expval{\phi^2}
    =\left[\left(\frac{N_c+N_f}{3N_c-N_f}\right)^{N_c-N_f}\frac{\Lambda^{3N_c-N_f}}{m^{N_c-N_f}}\right]^{\frac{1}{N_c}}
    \\
    &= \frac{2\mu^3}{g_t^2 m} \left(
    \frac{1}{3}N_c + \mathcal{O}(N_f) + \mathcal{O}(N_c^{-1})
    \right) e^{-8\pi^2/g_t^2},
\end{align}
which implies $f_\pi^2 \sim N_c$ for $N_c \gg 1$ and $N_f \ll N_c$. 
Therefore, the scaling of $f_\pi^2$ in ASQCD agrees with that of QCD and there is no evidence of the phase transition previously claimed in the literature. 

\subsection{The $\eta'$ Mass and the Topological Susceptibility}

The topological susceptibility $\chi_t$ is an important non-perturbative quantity in the QCD vacuum. It is given by 
\begin{align}
\chi_t\equiv\lim_{V\rightarrow\infty}\frac{\expval{Q_t^2}}{V}
= \left(\dod[2]{{\cal E}}{\theta}\right)_{\theta=0},
\end{align} 
where $Q_t$ is the integer valued topological charge of the gauge field in the four dimensional volume $V$, and ${\cal E}$ is the energy density of the vacuum. Note that it vanishes identically in the presence of massless quarks because $\theta$ can be rotated away by chiral symmetry. In \cite{Witten:1979vv}, the topological susceptibility in pure Yang--Mills theory was compared to the $\eta'$ mass in QCD with massless quarks using the large $N_c$ limit where fermion loops are suppressed by $1/N_c$ (which is actually $N_f/N_c$), and it was shown that up to corrections of order $1/N_c$ the mass of the $\eta'$ meson could be related to the topological susceptibility as\footnote{Note that \eqref{eq:wittenetaprime} differs from what appears in \cite{Witten:1979vv} by a factor of $2$. This is due to a difference in our definitions of $f_\pi$ by a factor of $\sqrt{2}$ as can be seen by calculating $\expval{0|\partial_\mu j_5^\mu|\eta'}=\sqrt{2N_f}f_\pi m_{\eta'}^2$ in our convention.}
\begin{align}
    f_\pi^2 m_{\eta'}^2 =& 2N_f\left(\dod[2]{{\cal E}}{\theta}\right)^\text{no quarks}_{\theta=0}. \label{eq:wittenetaprime}
\end{align}
We are in a position to confirm this relationship in ASQCD. We have already calculated $f_\pi^2$ in the appropriate limit above. For the topological susceptibility, consider perturbing the pure super-Yang-Mills theory with AMSB. Using the superpotential from gaugino condensation, $W=N_c\Lambda^3$, the vacuum energy is given via tree-level AMSB as
\begin{align}
    {\cal E} = -3mW+c.c. = & -3mN_c^2\mu^3g_t^{-2}\mathrm{e}^{-\frac{8\pi^2}{g_t^2}+i\frac{\theta}{N_c}} + c.c.,
    \\
    \left(\dod[2]{{\cal E}}{\theta}\right)^\text{no quarks}_{\theta=0} =& \frac{6m\Lambda^3}{N_c} = 6m\mu^3g_t^{-2}e^{-\frac{8\pi^2}{N_cg^2}}  .
\end{align}
Here we also show the expressions in terms of the 't Hooft coupling $g_t^2 = N_c g^2$ to show explicitly that $\mathcal{E}=\mathcal{O}(N_c^2)$. Finally, we use a result from \cref{sec:massspectrum} for $m_{\eta'}^2$ in ASQCD with $N_f<N_c$. Namely, defining $x=N_f/N_c$ we have\footnote{Note that the result quoted in \cite{Csaki:2023yas} was missing a factor of 2 that comes from properly integrating out the $\rho$ (see \cref{app:kinetic}). This factor of 2 does not change any of the conclusions in that work, but is important for us here.}
\begin{align}
    m_{\eta'}^2 =& \frac{2(3-x)^2x}{(1+x)(1-x)^2}m^2 = 18 m^2 x + \mathcal{O}(x^2). \label{eq:metap2}
\end{align}
In terms of $x$, we can compare this to the other two quantities in \cref{eq:wittenetaprime}
\begin{align}
    \left(\dod[2]{{\cal E}}{\theta}\right)^\text{no quarks}_{\theta=0} =& \frac{6m\Lambda^3}{N_c},
    &
    f_\pi^2 =& 2\Lambda^2 \left(\frac{1+x}{3-x}\frac{\Lambda}{m}\right)^{1-x},
\end{align}
and we can see that
\begin{align}
    f_\pi^2 m_{\eta'}^2
    &= 4 \frac{x(3-x)}{(1-x)^2} \left( \frac{3-x}{1+x} \frac{m}{\Lambda} \right)^x m \Lambda^3
    \stackrel{x\rightarrow 0}{\xrightarrow[\mspace{50mu}]{}} 12 x m\Lambda^3
    = 2N_f \left(\dod[2]{{\cal E}}{\theta}\right)^\text{no quarks}_{\theta=0}.
\end{align}
which to leading order in $1/N_c$ agrees perfectly with $2N_f$ times the topological susceptibility. Thus we have shown that \eqref{eq:wittenetaprime}, a relationship derived in the context of large $N_c$ QCD, holds also in large $N_c$ ASQCD. Away from the large $N_c$ limit, the factor $(\Lambda/m)^{x}$ accounts for the difference in the beta function and hence the dynamical scales between $N_f = 0$ (SYM) and $N_f \neq 0$. The remaining factor $((3-x)/(1+x))^{1-x}$ is the impact of the mixing between $gg$ and $q\bar{q}$ configurations, and this is the first time this has been presented in the literature to the best of our knowledge.

In the presence of a small but finite quark mass $m_q < m$, the topological susceptibility does not vanish even with dynamical quarks. In this case, we can compare the topological susceptibility and the mass of $\pi$ within the same theory. Assuming $N_f$ flavor of quarks with the same mass $m_q$ for simplicity, we obtain\footnote{Note that the explicit calculation for $N_f=3,\,N_c=4$ in \cref{sec:Nf3Nc4} agrees with this expression.}
\begin{align}
    m_\pi^2 &= \frac{2(7N_c - N_f)}{N_c+N_f} m_q m
    = \frac{2(7 - x)}{1+x} m_q m,
\end{align}
and
\begin{align}
    \left(\dod[2]{{\cal E}}{\theta}\right)^\text{with quarks}_{\theta=0} 
    &= 2 m_q \Lambda^3 \frac{1}{N_f} \frac{7-x}{(3-x)}\left(\frac{3-x}{1+x} \frac{m}{\Lambda} \right)^{x} .
\end{align}
This should be compared to
\begin{align}
    f_\pi^2 m_\pi^2
    &= 4 m_q \Lambda^3 \frac{7-x}{(3-x)}\left(\frac{3-x}{1+x} \frac{m}{\Lambda} \right)^{x}
    = 2 N_f \chi_t .
\end{align}
They agree completely including the $N_f$ dependence. This confirms that the relation found in \cite{Veneziano:1979ec} for QCD also holds in ASQCD, a result further supported by the lattice result in~\cite{Dimopoulos:2018xkm}.

\subsection{New Scaling Predictions for $f_\pi^2$}

\begin{figure}
    \centering
    \includegraphics[width=0.8\linewidth]{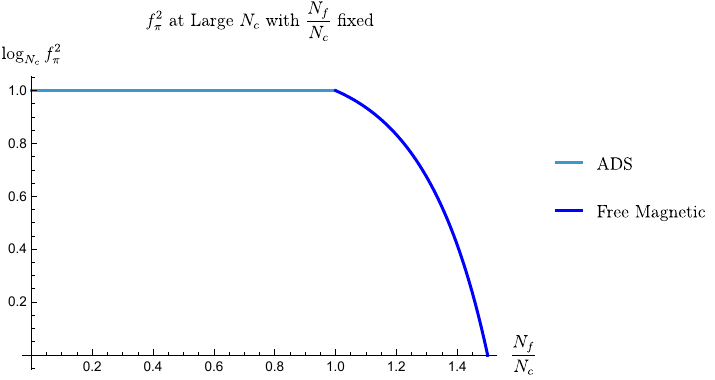}
    \caption{The dependence of $\log_{N_c}(f_\pi^2)$ on $N_f/N_c$ is shown, in the limit $N_c\to\infty$ with $N_f/N_c$ held fixed, for $N_f<3N_c/2$. We use~\cref{eq:ADSVEV,eq:FreeMagVEV}. For $N_f<N_c$ the expected scaling behavior $f_\pi^2\propto N_c$ as predicted by 't Hooft diagram approach is observed. In contrast, for $N_f>N_c$ the scaling is suppressed. Although this plot is constructed in the $N_c\to\infty$ limit, our results can also be used to predict the dependence on $N_f/N_c$ at finite $N_c$. }
    \label{fig:fpi}
\end{figure}

\begin{figure}
    \centering
    \includegraphics[width=0.6\linewidth]{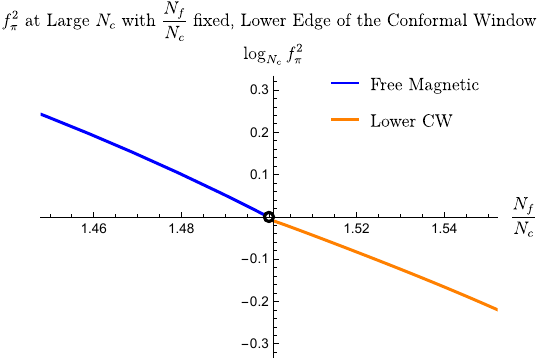}
    \caption{The $N_f/N_c$ dependence of $\log_{N_c}(f_\pi^2)$ is shown in the $N_c\to\infty$ limit with $N_f/N_c$ held fixed, near the lower edge of the Conformal Window. We use~\cref{eq:FreeMagVEV,eq:LCWVEV}. The point at exactly $N_f=3N_c/2$ is omitted, reflecting the fact that AMSB cannot be applied to obtain the non-SUSY limit due to the classical conformal invariance there. An interpolation between the chiral symmetry breaking results for $N_f<3N_c/2$ and $N_f>3N_c/2$ is expected.}
    \label{fig:fpicwL}
\end{figure}

\begin{figure}
    \centering
    \includegraphics[width=0.8\linewidth]{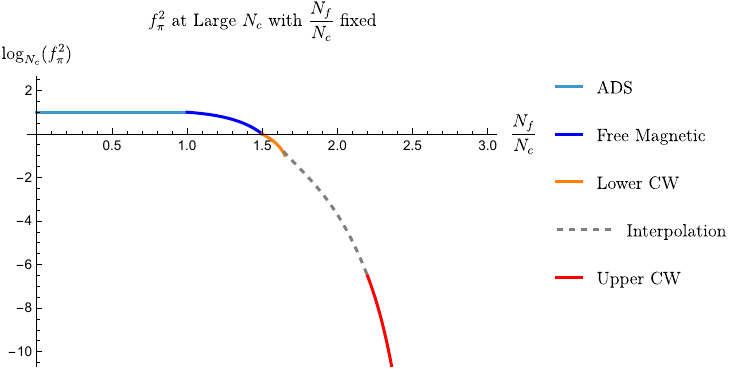}
    \caption{The $N_f/N_c$ dependence of $\log_{N_c}(f_\pi^2)$ is shown in the $N_c\to\infty$ limit with $N_f/N_c$ held fixed for the full range $0<N_f/N_c < 3$. We use~\cref{eq:ADSVEV,eq:FreeMagVEV,eq:LCWVEV,eq:UCWVEV}. In the conformal window, results from perturbing around the electric and magnetic Banks-Zaks fixed points are shown in red and orange, respectively, and a smooth interpolation between them is shown in dashed gray. Note that $f_\pi^2 \to 0$ as $N_f\to 3N_c$ which is because the AMSB becomes ineffective when the SUSY limit is classically conformal.}
    \label{fig:fpifull}
\end{figure}

We can go further, since we have results for the minima when $N_f\geq N_c$, allowing us to take the limit $N_c\to \infty$ with $N_f/N_c$ held fixed. This kind of prediction has never been worked out to the best of our knowledge. This corresponds to higher-genus corrections to the dominant planar diagrams with many holes \cite{tHooft:1973alw,Veneziano:1976wm}. One fermion loop contribution differs by a factor $1/N_c$, but comes with a factor of the number of fermions $N_f$. When $N_f$ is small, the planar diagrams without holes dominate. While as we increase the number of fermions, different behavior is observed. Our results are summarized in  \cref{fig:fpi,fig:fpicwL,fig:fpifull} and the functional form of the scaling is given in \cref{tab:fpiscaling}.

\begin{table}[]
    \centering
    \renewcommand{\arraystretch}{2.0}
    \begin{tabular}{|c||c|c|c|c|}
         \hline & $0<x\leq 1$ & $1\leq x < 3/2$ & $3/2 \lesssim x \ll 3$ & $3/2 \ll x \lesssim 3$ \\ \hline\hline
         $f_\pi^2\propto\;$& $N_c \left(\frac{3+3x}{3-x}\right)^{1-x}$ 
         &
         $N_c^{\frac{6-4x}{6-5x+x^2}} \left(\frac{3-2x}{3}\right)^{\frac{2(x-1)}{2-x}}$
         & 
         $\frac{14}{3}(x-\frac{3}{2})^2 N_c^{-4(x-\frac{3}{2})}$
         &
         $x^{-1}N_c^{-\frac{x^2-2x+3}{(x-3)^2}}$
         \\\hline
    \end{tabular}
    \caption{Scaling of $f_\pi^2$ in the limit $N_c\to\infty$ with $x\equiv N_f/N_c$ held fixed. Note that in the $N_c\to\infty$ limit the Quantum-Modified and s-confining regimes can be taken to agree with the Free-Magnetic regime. For the Conformal Window, we only have results perturbatively around the Banks-Zaks fixed points at either edge. In lower edge of the Conformal Window, in particular, we show the leading order behavior as $x\to 3/2$.}
    \label{tab:fpiscaling}
\end{table}

\subsection{New Scaling Predictions for $\expval{GG}$}
Just as with the chiral condensate, 't Hooft's planar diagrams suggest a scaling behavior for $\expval{GG}$ in the large $N_c$ limit. Specifically, we expect $\expval{GG}\propto\, N_c^2$, assuming the canonical normalization for the gluon field. We will demonstrate how they can be computed in the next section. Here we present only the results. As we have explicit results for gluon condensates in ASQCD, we can both confirm this scaling as well as predict new scaling when the ratio $N_f/N_c$ is held fixed. 

\begin{figure}
    \centering
    \includegraphics[width=0.8\linewidth]{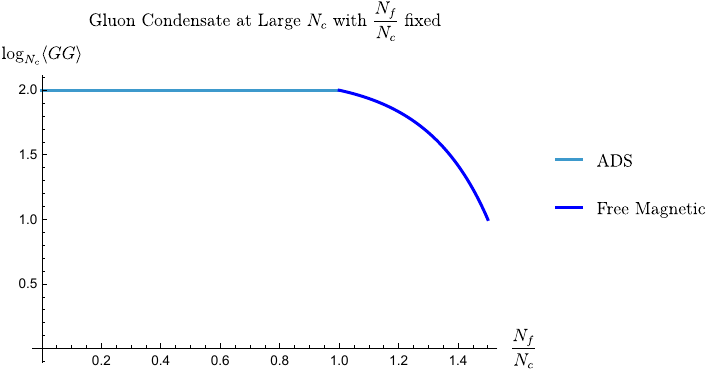}
    
    \caption{The scaling of the gluon condensate in the $N_c\to\infty$ limit with $N_f/N_c$ held fixed is shown. We have used~\cref{eq:ADSgluoncondense,eq:Freemaggluoncondense} The expected $N_c^2$ scaling is observed when $N_f < N_c$, whereas the power law decreases when $N_f>N_c$. The region $0<N_f/N_c<1.5$ is shown here.}
    \label{fig:GGlowx}
\end{figure}

\begin{figure}
    \centering
    \includegraphics[width=0.8\linewidth]{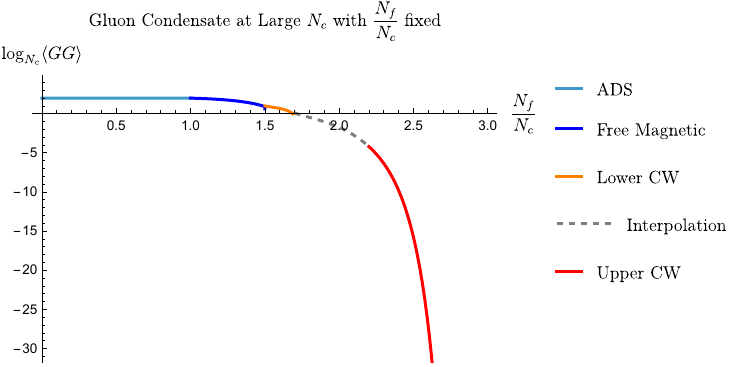}
    \caption{The scaling behavior of the gluon condensate in the $N_c\to\infty$ limit with $N_f/N_c$ held fixed is shown. We use~\cref{eq:ADSgluoncondense,eq:Freemaggluoncondense,eq:LCWgluoncondense,eq:UCWgluoncondense} Here we show the full range $0<N_f/N_c<3$. Note that the results from the upper and lower edges of the window are interpolated.}
    \label{fig:GGfull}
\end{figure}

\begin{figure}
    \centering
    \includegraphics[width=0.8\linewidth]{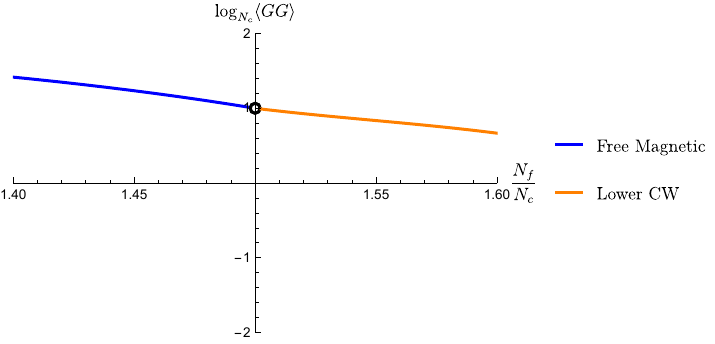}
    \caption{The scaling behavior of the gluon condensate in the $N_c\to\infty$ limit with $N_f/N_c$ held fixed is shown. We use~\cref{eq:LCWgluoncondense,eq:UCWgluoncondense}. Here we focus on the region around $\frac{N_f}{N_c}\sim 1.5$, which corresponds to the lower edge of the broken conformal window. Note that the point at exactly $\frac{N_f}{N_c}=1.5$, there are no results from AMSB due to classical conformal invariance, and so results from above and below are interpolated.}
    \label{fig:GGboundary}
\end{figure}

In \Cref{fig:GGlowx,fig:GGfull,fig:GGboundary}, using the results~\cref{eq:Gluecondensebasics,eq:Gluinocondensebasics}, we show the power law $\log_{N_c}\expval{GG}$ in the $N_c\to \infty$ limit with $N_f/N_c$ held fixed, in the different regions of interest. Completely analogous to our results for $f_\pi^2$, we find that as $N_f>N_c$ the power law decreases at an accelerating rate. We note that the factor $\log \Lambda/\mu$ is logarithmically suppressed in powers of $N_c$, and thus becomes negligible in the $N_c\rightarrow\infty$ limit.

\subsection{Derivation of Wess-Zumino-Witten (WZW) term}


We can derive the WZW term based on the technique in \cite{DHoker:1984izu} by integrating out mesino (fermion) for the ADS case. We will consider a situation after condensation $\expval{Q^a_i}=\expval{\tilde{Q}^a_i}=v\delta^a_i$, with $\xi^2=U$ ($U$ is a $N_f\times N_f$ unitary matrix in~\cref{subsec:chiralLagADS}). 
\begin{align}\label{eq:VEV}
Q&=v\xi+\theta \psi_{Q}+\theta^2 F, \qquad
\tilde{Q}^T=v\xi +\theta \psi_{\tilde{Q} }+\theta^2 \tilde{F}.
\end{align}
Note that the VEV \cref{eq:VEV} is diagonal in color and flavor indices. It seems that gluino $\lambda$ is unrelated to flavor, but because of this color-flavor diagonal vacuum, gluino $\lambda$ is related to the flavor.

Let us consider an effective action
\begin{align}
W=-i\log\left[ \int dQd\tilde{Q} dV \exp\left[ i\int d^4x \mathcal{L} \right] \right].
\end{align}
When we transform fermion by $Q\rightarrow \xi^*Q,\ \tilde{Q}\rightarrow \xi^\dag \tilde{Q}$ in order to diagonalize the mass matrix of quarks and gauginos, additional terms arise from the Jacobian in the path integral. The Jacobian appears in the following way
\begin{align}
&D(\xi^*Q)D(\tilde{Q}\xi^\dag)\rightarrow DQ J(\xi^T) D\tilde{Q}J(\xi),\\
&J(\xi)=\exp[iN_c\int \operatorname{Tr}(\xi^\dag d\xi)^5], \ 
J(\xi^T) =\exp\left[-iN_c\int \operatorname{Tr}(\xi d\xi^\dag)^5\right],
\end{align}
where we have used the invariance of trace under transpose and cyclic permutation. We have additional terms in the effective action as
\begin{align}
    W'&=W-i\log J(\xi)-i\log J(\xi^T)\nonumber\\
    &\supset N_c\int \operatorname{Tr}\left[(\xi^\dag d\xi)^5-(\xi d\xi^\dag)^5\right]
   =N_c\int \operatorname{Tr}[(U^\dag dU)^5],
\end{align}
where we have shown the calculation in the last line up to local counter terms in \cref{sec:WZWtracecalc}. Therefore, we can derive the WZW term from the calculation of the mesino loop. Note that the factor $N_c$ comes from the possible mass terms invariant under transformation, which is discussed in~\cref{sec:originNc}.

\section{Calculation of Condensates}\label{sec:condensate}

In perturbation theory, we cannot compute non-perturbative condensates of composite operators. Fortunately they are computable with our method. We show how they can be computed, and compare their scaling with expectations from dimensional analysis on the QCD side. Since we only have access to these scaling laws in the asymptotic regions $m\ll \Lambda$ and $m\gg \Lambda$, we cannot rigorously address whether the intermediate $m\sim\Lambda$ region smoothly interpolates between the two. Nevertheless, we find that the ratios of condensates within a supermultiplet are consistent between the near-SUSY and non-SUSY limits. We take this as yet another piece of evidence suggesting that the two limits are continuously connected.

\subsection{The Expectations from QCD}

For $m\gg \Lambda$, we expect that the quark bilinear condensate should scale as $\Lambda_\text{QCD}^3$ on the grounds of dimensional analysis\footnote{Dimensional analysis is of course incapable of determining the coefficients of such relations, so we can only compare scaling. }. Matching above and below the SUSY-breaking scale should relate the dynamical scales by one-loop matching valid for $m \gg \Lambda$,
    \begin{align}
        \Lambda_\text{QCD}^{\frac{11}{3}N_c-\frac{2}{3}N_f} =&\, m^{\frac{2}{3}N_c+\frac{1}{3}N_f}\Lambda_\text{SUSY}^{3N_c-N_f}.
        \label{eq:scalematch}
    \end{align}
We use this matching to express the ASQCD results with $\Lambda_\text{QCD}$ held fixed. We similarly have the expectation  that for $m\gg\Lambda$ the gluon condensate should scale as $\Lambda_\text{QCD}^4$. In the $m\gg \Lambda$ limit, but still without integrating out the superpartners, the presence of these QCD condensates can act as a source for the superpartner condensates, as shown in \cref{fig:factorm}. The presence of a mass insertion on the internal gluino line leads to the expectation that the condensates will differ in scaling by one factor of $m$:
\begin{align}
    \expval{\tilde{q}^*\tilde{q}}\propto \frac{\Lambda_{QCD}^3}{m}\\
    \expval{qq}\propto\Lambda_{QCD}^3\\
\expval{\lambda\lambda}\propto\frac{\Lambda_{QCD}^4}{m}\\
\expval{GG}\propto\Lambda_{QCD}^4.
\end{align}
As expected, as $m\rightarrow\infty$, the superpartner condensates decrease, corresponding to their decoupling as the non-SUSY limit is approached.
In terms of the SUSY dynamical scale instead the expectation is:
	\begin{align}
		\langle\tilde{q}^*\tilde{q}\rangle \propto\,& \Lambda^2 \left(\frac{m}{\Lambda}\right)^{\frac{5(N_f-N_c)}{11N_c-2N_f}} 
		\\
		\langle\bar{q}q\rangle \propto\,& m\Lambda^2 \left(\frac{m}{\Lambda}\right)^{\frac{5(N_f-N_c)}{11N_c-2N_f}} 
	\\
        \langle GG\rangle \propto&\, m\Lambda^3 \left(\frac{m}{\Lambda}\right)^{\frac{3(N_c-2N_f)}{11N_c-2N_f}}
        \\
        \langle \lambda\lambda\rangle \propto&\, \Lambda^3 \left(\frac{m}{\Lambda}\right)^{\frac{3(N_c-2N_f)}{11N_c-2N_f}}
    \end{align}
\begin{figure}
    \centering
    \begin{tikzpicture}[scale=1.2]
    \begin{feynman}
        \node[crossed dot,label=left:$\expval{\tilde{q}^*\tilde{q}}$] (i1) at (-1,0);
        \node[crossed dot,label=right:$\expval{\bar{q}q}$] (i2) at (1,0);
        \vertex (m1) at (0,1);
        \vertex (m2) at (0,-1);

        \diagram*{
        (i1) -- [charged scalar, quarter left] (m1),
        (i1) -- [charged scalar, quarter right] (m2),
        (m2) -- [fermion, quarter right] (i2),
        (m1) -- [fermion, quarter left] (i2),
        (m1) -- [boson, anti majorana, insertion=0.5] (m2),
        };
    \end{feynman}
\end{tikzpicture}
\hspace{0.2in}
\begin{tikzpicture}[scale=1.2]
    \begin{feynman}
        \node[crossed dot,label=left:$\expval{GG}$] (i1) at (-1,0);
        \node[crossed dot,label=right:$\expval{\lambda\lambda}$] (i2) at (1,0);
        \vertex (m1) at (0,1);
        \vertex (m2) at (0,-1);

        \diagram*{
        (i1) -- [gluon, quarter left] (m1),
        (i1) -- [gluon, quarter right] (m2),
        (m2) -- [boson,  fermion, quarter right] (i2),
        (m1) -- [boson,  fermion, quarter left] (i2),
        (m1) -- [boson, anti majorana,insertion=0.5] (m2),
        };
    \end{feynman}
\end{tikzpicture}
    \caption{In the $m\gg \Lambda$ limit the QCD quark and gluon condensates can act as sources for the superpartner condensates. The leading-order diagrams are shown here. Note the mass-insertion on the central gaugino line.}
    \label{fig:factorm}
\end{figure}
Below, we find the same factor of $m$ scaling difference when calculating the condensates in the $m\ll\Lambda$ limit. The only exceptions are in the Quantum-Modified phase, where the coefficient of the leading-order result vanishes, and in the conformal window where the scaling results are only confirmed to leading order in the perturbative expansion around the BZ fixed points.

\subsection{Method for ASQCD}
Wherever we have identified a $\chi$SB minimum we can explicitly obtain the expectation value of the meson operator $\expval{M}$. As a chiral superfield
\begin{align}
    \expval{M} =& \expval{\tilde{q}^*\tilde{q}}
    +\theta^2 \expval{\bar q q} .
\end{align}
We get the lowest component when identifying the vacuum by minimizing the scalar potential (see \Cref{sec:vacua}). The F-component can be obtained by the equations of motion. Generically the $F$-term part of the Lagrangian in AMSB in canonical normalization is
\begin{align}
    \mathcal{L} \supset& F_{\tilde{X}}\dpd{W}{\tilde{X}} +h.c. + F_{\tilde{X}}F_{\tilde{X}}^\dagger + \frac{1}{c_X Z_X} m(1-\frac{\gamma_X}{2})(F_{\tilde{X}}\tilde{X}^\dagger + h.c.).
\end{align}
Up to an overall dimensionful scale, then, we will have
\begin{align}
    \expval{\bar{q}q} =& -\expval{
    \dpd{W}{\tilde{M}}
    }^\dagger - m\expval{
    (1-\frac{\gamma_M}{2}) \frac{\tilde{M}}{Z_M}
    },
\end{align}
where here this is generically a matrix expression. It is sometimes more convenient to calculate in terms of the non-canonically normalized $M=\tilde{M}/\sqrt{ Z_M}$: 
\begin{align}
    -\expval{\frac{1}{\sqrt{ Z_M}}\dpd{W}{M}}^\dagger-m\expval{(1-\frac{\gamma_M}{2}) \frac{M}{\sqrt{  Z_M}}}.
\end{align}
Here, the derivative is well-defined, and the expectation values should evaluated at $\mu=\phi_\text{min, can}$, where the subscript denotes the field value at the potential minimum, expressed in the canonical normalization of the matter field. 

In this way, we can obtain the fermion and sfermion bilinear condensates from the existing calculations of the vacua.

Now we will derive how to calculate the gluon and gluino condensates. First, the Lagrangian for the vector multiplet is
\begin{align}
    \mathcal{L}_\text{vec}
    &=\frac{1}{16\pi i}\int d^2\theta \,\tau W_\alpha W^\alpha +\text{ h.c.}
\end{align}
Here, we regard the holomorphic coupling as spurion superfield
\begin{align}
    \tau =&\,\tau + \sqrt{2}\psi_\tau\theta + \mathcal{F}_\tau \theta^2 .
\end{align}
Throughout the following analysis we set $\psi_\tau=0$, as it does not play a significant role. The scalar and F-components of this spurion then act as sources for the gluon and gluino condensates:
\begin{align}
    \mathcal{L}_\text{vec} \supset& \frac{1}{16\pi i} \tau (GG+\bar{\lambda} i \slashed{D}\lambda) + \frac{1}{16\pi i} \mathcal{F}_\tau \lambda\lambda  .
\end{align}
Take the path integral to be
\begin{align}
    Z=\int D[\text{fields}] \exp{i\int d^4x \left(\int d^4\theta K + \int d^2\theta W + \int d^2\bar\theta W^*\right)} .
\end{align}
As a function of the spurion superfield $\tau$, the superpotential can be expanded as following
\begin{align}
 \int d^2\theta\ W(\tau+\mathcal{F}_\tau\theta^2)&= \int d^2\theta\ W(\tau) + \mathcal{F}_\tau \dpd{W}{\tau} .
\end{align}
Treating $\tau$ and $\mathcal{F}_\tau$ as sources, we can calculate the condensates via the path integral like so 
\begin{align}
16\pi i\frac{\partial\ln Z}{\partial\tau}
&=\expval{GG+\bar{\lambda}i\slashed{D}\lambda}=\expval{GG},\\
16\pi i\frac{\partial\ln Z}{\partial\mathcal{F}_\tau}
&=\expval{\lambda\lambda}.
\end{align}
In the case of ASQCD we can directly determine the minima of the theory, and thus evaluate the path integral explicitly. We can then use our determination of the minima to calculate the gluon and gluino condensates like so
\begin{align}
    \expval{GG} =& 16\pi i \frac{\partial}{\partial\tau}\ln\mathcal{Z} = 16\pi i \frac{\partial \expval{V}}{\partial \tau} ,
    \\
    \expval{\lambda\lambda} =& 16\pi i \frac{\partial}{\partial F_\tau}\ln \mathcal{Z} = 16\pi i \frac{\partial \expval{W}}{\partial\tau} ,
\end{align}
where $\expval{V}$ is the vacuum energy and $\expval{W}$ is the expectation value of the scalar component of the superpotential. The dynamical scale is given by the holomorphic gauge coupling
\begin{align}
\Lambda =& \mu\, e^{\frac{2\pi i}{3N_c-N_f}\tau}\, ,
\end{align}
where $\mu$ is the UV cutoff scale. We then have
\begin{align}
    \expval{GG} =& -\frac{32\pi^2}{3N_c-N_f} \Lambda \frac{\partial\expval{V}}{\partial \Lambda}\ ,
    \\
    \expval{\lambda\lambda} =& -\frac{32\pi^2}{3N_c-N_f} \Lambda \frac{\partial \expval{W}}{\partial \Lambda}\ .
\end{align}
But note that these are not canonically normalized (with respect to the gauge and gaugino kinetic terms). To normalize them properly, we must multiply by $\tau/4\pi i$ such that
\begin{align}
    \expval{GG}_c =& 4 \ln\frac{\Lambda}{\mu} \Lambda \dpd{\expval{V}}{\Lambda}\ ,\label{eq:Gluecondensebasics}
    \\
    \expval{\lambda\lambda}_c =&  4 \ln\frac{\Lambda}{\mu} \Lambda \dpd{\expval{W}}{\Lambda}\ .\label{eq:Gluinocondensebasics}
\end{align}
In what follows we will drop the subscript and use $\expval{GG}$ and $\expval{\lambda\lambda}$ to refer to the canonically normalized condensates. When identifying power-law scaling behavior, we neglect the $\ln(\Lambda/\mu)$ factor.

We will take conventions such that the moduli are real for simplicity. We can now go through each phase and use the minima calculated previously to determine the condensates in the way specified above. For a summary of the leading power-law scaling, see \cref{tab:condensate-scaling}. The calculations leading to these results are detailed in the rest of the section.

\begin{table}[htb]
    \centering
    \renewcommand{\arraystretch}{1.2}
    {
\begin{tabular}{|c||c|c|c|c|}
     \hline
     Phase & $\expval{\tilde{q}^*\tilde{q}}$ & $\expval{\bar{q}q}$ & $\expval{\lambda\lambda}$ & $\expval{GG}$ \\
     \hline\hline
     ADS & $m^{-1}\Lambda^3 (m/\Lambda)^{\frac{N_f}{N_c}}$ & $\Lambda^3 (m/\Lambda)^{\frac{N_f}{N_c}}$ & $\Lambda^3 (m/\Lambda)^{\frac{N_f}{N_c}}$ & $m\Lambda^3 (m/\Lambda)^{\frac{N_f}{N_c}}$
     \\\hline
     QM & $\Lambda^2$ & $m\Lambda^2$ &
     {$m^5\Lambda^{-2}$} & $m^2\Lambda^2$
     \\\hline
     s-conf & $\Lambda^2(m/\Lambda)^{\frac{1}{N_f-2}}$ & $m\Lambda^2(m/\Lambda)^{\frac{1}{N_f-2}}$ &  $\Lambda^3(m/\Lambda)^{\frac{N_f}{N_f-2}}$ & $m\Lambda^3(m/\Lambda)^{\frac{N_f}{N_f-2}}$
     \\\hline
     FM & $\Lambda^2 (m/\Lambda)^{\frac{N_f-N_c}{2N_c-N_f}}$ & $m\Lambda^2 (m/\Lambda)^{\frac{N_f-N_c}{2N_c-N_f}}$ & $\Lambda^3 (m/\Lambda)^{\frac{N_f}{2N_c-N_f}}$ & $m\Lambda^3 (m/\Lambda)^{\frac{N_f}{2N_c-N_f}}$
     \\\hline
     LCW & $\Lambda^2 (m/\Lambda)^{\frac{2}{2-\alpha}}$ & $\Lambda^3 (m/\Lambda)^{\frac{4}{2-\alpha}}$ & $\Lambda^3 (m/\Lambda)^{\frac{6}{2-\alpha}}$ & $\Lambda^4 (m/\Lambda)^{\frac{8}{2-\alpha}}$
     \\\hline
     UCW & $\Lambda^2 (m/\Lambda)^{\frac{1}{\epsilon}}$ & $m\Lambda^2(m/\Lambda)^{\frac{1}{\epsilon}}$ & $m \Lambda^2 (m/\Lambda)^{\frac{2}{\epsilon}}$ & $m^2 \Lambda^2 (m/\Lambda)^{\frac{2}{\epsilon}}$
     \\
     \hline
\end{tabular}
}
    \caption{Leading power-law scaling of the various condensates with respect to the SUSY-breaking scale $m$ and the holomorphic scale $\Lambda$. Here, the phases refer to Affleck--Dine--Seiberg (ADS), quantum modified moduli space (QM), s-confinement, free magnetic (FM), lower conformal window (LCW), and upper conformal window (UCW).}
    \label{tab:condensate-scaling}
\end{table}

\subsection{ADS Phase }
Because $M=\tilde{Q}Q$ we have that $F_M = 2\expval{\phi F_\phi}$ and thus from our existing results we can calculate
\begin{align}
    \expval{\bar q q} =& 
    2\left(
    1-\frac{4N_f(N_c+N_f)}{3N_c-N_f}
    \right)
    \Lambda^3
    \left(
    \frac{3N_c-N_f}{4N_f(N_c+N_f)}
    \frac{m}{\Lambda}
    \right)^{\frac{N_f}{N_c}} .
\end{align}
Note that
\begin{align}
    \expval{\tilde{q}^*\tilde{q}} \propto&\quad m^{\frac{N_f}{N_c}-1} \Lambda^{3-\frac{N_f}{N_c}}, \label{eq:ADSsquarkcondense}\\
    \expval{\bar q q} \propto&\quad m^{\frac{N_f}{N_c}} \Lambda^{3-\frac{N_f}{N_c}},\label{eq:ADSquarkcondense}
\end{align}
which differ by a factor of $m$ in accordance with our expectations.

Given $V,W,$ and $\expval{\phi}$,  the gluon and gluino condensates can be computed as follows:
\begin{align}
    \expval{GG} =& -2\frac{(3N_c-N_f)^2(4N_c+N_f-1)}{N_c(N_c+N_f)}  \left(\frac{3N_c-N_f}{4N_f(N_c+N_f)}\frac{m}{\Lambda}\right)^{\frac{N_f}{N_c}}m\Lambda^3\ln\frac{\Lambda}{\mu}\ ,
    \\
    \expval{\lambda\lambda} =& 
    \frac{4(3N_c-N_f)(N_c-N_f)}{N_c}
    \left(
    \frac{3N_c-N_f}{4N_f(N_c+N_f)}
    \frac{m}{\Lambda}
    \right)^{\frac{N_f}{N_c}}
    \Lambda^3 \ln\frac{\Lambda}{\mu}\ .
\end{align}
Note that to leading order in $m$:
\begin{align}
    \expval{\lambda\lambda}\propto&\ 
    m^{\frac{N_f}{N_c}} \Lambda^{3-\frac{N_f}{N_c}},\label{eq:ADSgluinocondense}\\
    \expval{GG}\propto&\  m^{1+\frac{N_f}{N_c}}\Lambda^{3-\frac{N_f}{N_c}} \label{eq:ADSgluoncondense},
\end{align}
which again differ by a factor of $m$, conforming with expectations.

\begin{figure}[htb] 
    \centering
    \includegraphics[width=\linewidth]{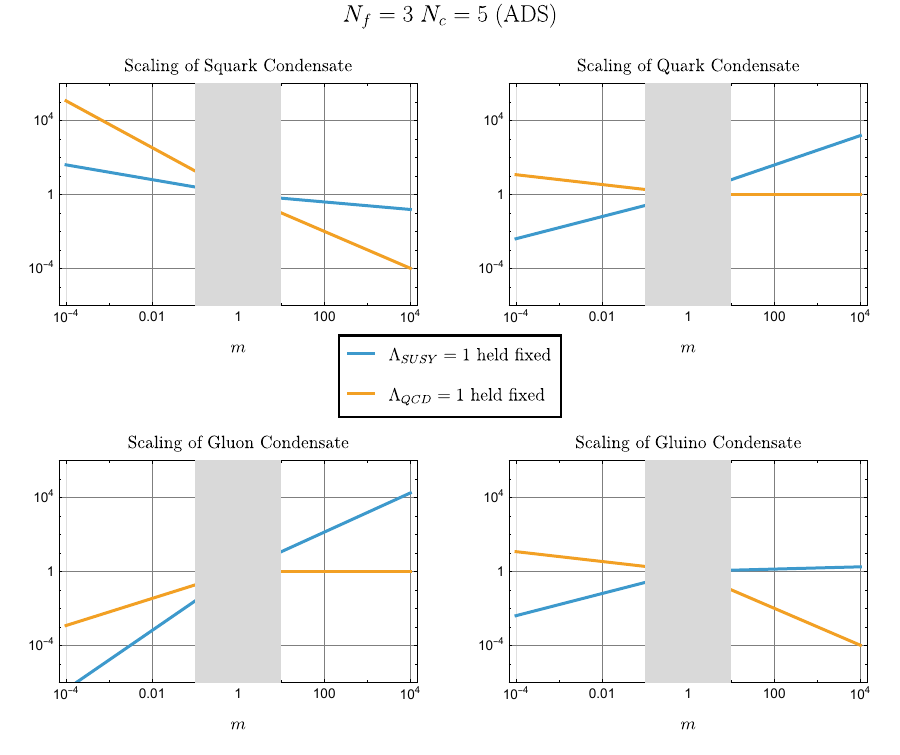}
    \caption{The scaling of the various bilinear condensates in the AMSB-perturbed ADS phase is shown as a power law in $m$, based on~\cref{eq:ADSsquarkcondense,eq:ADSquarkcondense,eq:ADSgluoncondense,eq:ADSgluinocondense} in units with $\Lambda_{SUSY}=1$ held fixed (blue) and $\Lambda_{QCD}=1$ held fixed (orange). The latter corresponds to exchanging $\Lambda_{SUSY}$ for $\Lambda_{QCD}$ via \cref{eq:scalematch}, which for this particular plot is $\Lambda_{SUSY} = \left(\Lambda_{QCD}^{49}/m^{13}\right)^{1/36}$.  The choice $N_f=3,N_c=5$ serves as a representative example; other values do not qualitatively alter the plots (so long as $0<N_f<N_c$). The gray region around $m\sim 1$ indicates the incalculable regime. The results in the $m\ll \Lambda$ region are \textit{exact} (in the ADS phase there is no ambiguity even of $\mathcal{O}(1)$ factors) coming from the AMSB calculation. The results in the $m\gg\Lambda$ region are, up to an unknown overall factor, uniquely determined by dimensional analysis (QCD only has one scale), scale matching as in \cref{eq:scalematch}, and the two-loop generation of superpartner condensates in \cref{fig:factorm} (which should be very accurate in the $m\gg\Lambda$ limit, and only inherits the uncertainty in the overall factor of the QCD condensates).}
    \label{fig:condensatesADS}
\end{figure}

See \cref{fig:condensatesADS} for a comparison of the scaling in the small and large SUSY-breaking regimes in terms of both $\Lambda_\text{SUSY}$ and $\Lambda_\text{QCD}$.

\subsection{Quantum Modified Moduli Space }
The properly normalized (that is, corresponding to the vev of the dimension 2 meson field) squark condensate obtained from the minimum is given at leading order by 
\begin{align}
    \expval{\tilde q^*\tilde q} =& \Lambda^2.\label{eq:QMsquarkcondese}
\end{align}
In terms of the canonically normalized $\phi$, we can obtain the quark condensate via 
\begin{align}
    \frac{1}{\Lambda}\expval{\bar q q} =& -\expval{\frac{\partial W}{\partial M}} - m \expval{M}, 
\end{align}
and hence
\begin{align}
        \expval{\bar q q} =& - m\Lambda^2.\label{eq:QMquarkcondense}
\end{align}
For the gluon condensate, we calculate at leading order
\begin{align}
     \expval{GG} =& -8N_c  m^2 \Lambda^2 \ln\frac{\Lambda}{\mu}.\label{eq:QMgluoncondense}
\end{align}
For the gluino condensate, using the leading order minima gives $\expval{W}=0$. However, expanding to higher order in $m/\Lambda$ we find
\begin{align}
    \expval{W} =&
    \frac{m^3}{\lambda^2} + \frac{3N_c-1}{2N_c \lambda^4}\frac{m^5}{\Lambda^2} \ ,
\end{align}
such that
\begin{align}
    \expval{\lambda\lambda} =& 4\frac{3N_c-1}{\lambda^4 N_c}  \frac{m^5}{\Lambda^2}\ln\frac{\Lambda}{\mu}\ .\label{eq:QMgluinocondense}
\end{align}
Note that the ADS and Free Magnetic results both have $\expval{\lambda\lambda}\propto (N_f-N_c)$, consistent with a leading-order cancellation when $N_f=N_c$.

It is important to note that the stability of this minimum is ambiguous, due to unknown coefficients in the K\"ahler potential \cite{Csaki:2022cyg}. Moreover, this minimum is not exact but rather the leading order result in powers of $m$. The results we have for this phase should be viewed with caution, but are bolstered if one assumes a degree of continuity between the ADS phase below and the $s$-confining phase above. Moreover we have checked that the inclusion of higher order K\"ahler terms -- non-negligible when field values are order $\Lambda$ -- only changes the coefficients of the sub-leading terms of the minimum and thus does not affect this leading-order cancellation. 

\begin{figure}[htb]
    \centering
    \includegraphics[width=\linewidth]{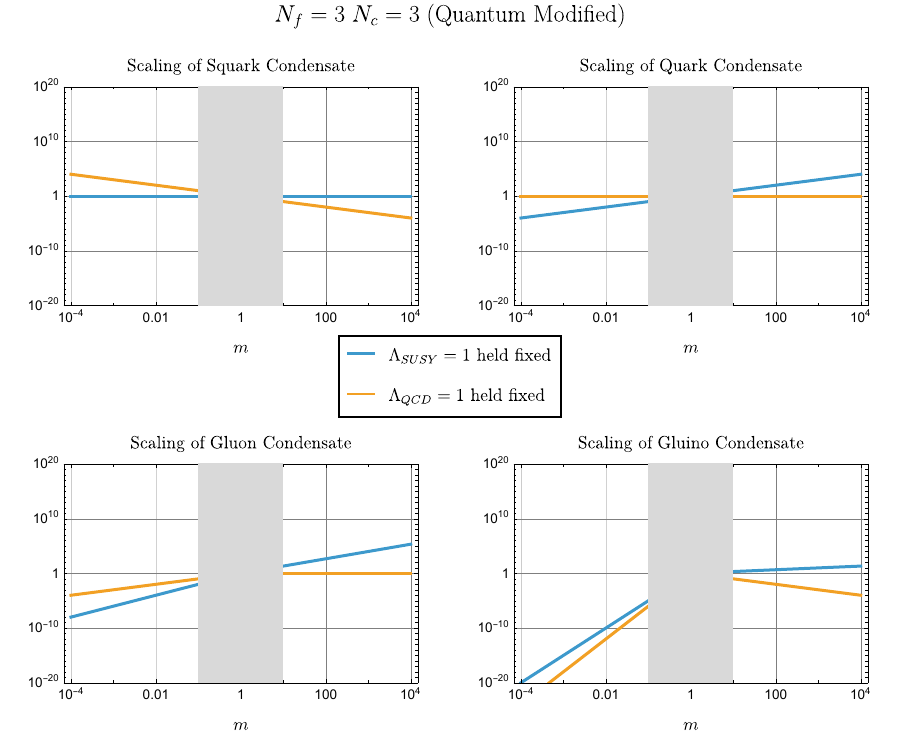}
    \caption{The scaling of the various bilinear condensates in the AMSB-perturbed Quantum-Modified phase is shown as a power law in $m$ based on~\cref{eq:QMsquarkcondese,eq:QMquarkcondense,eq:QMgluoncondense,eq:QMgluinocondense} in units with $\Lambda_{SUSY}=1$ held fixed (blue) and $\Lambda_{QCD}=1$ held fixed (orange). The latter corresponds to exchanging $\Lambda_{SUSY}$ for $\Lambda_{QCD}$ via \cref{eq:scalematch} which for this particular plot is $\Lambda_{SUSY} = (\Lambda_{QCD}^3/m)^{1/2}$.  The choice $N_f=3,N_c=3$ serves as a representative example; other values do not qualitatively alter the plots (so long as $N_f=N_c$). The gray region around $m\sim 1$ reflects the incalculability in that region. Note that the gluino condensate appears to be a broken line because of peculiar cancellation for this case as discussed in the text. A higher-order calculation would be interesting, but is difficult to carry-out given the lack of control over the K\"ahler potential. The results in the $m\ll \Lambda$ region are \textit{exact} (up to unknown $\mathcal{O}(1)$ factors not relevant for the scaling behavior) coming from the AMSB calculation. The results in the $m\gg\Lambda$ region are, up to an unknown overall factor, uniquely determined by dimensional analysis (QCD only has one scale), scale matching as in \cref{eq:scalematch}, and the two-loop generation of superpartner condensates in \cref{fig:factorm} (which should be very accurate in the $m\gg\Lambda$ limit, and only inherits the uncertainty in the overall factor of the QCD condensates).}
    \label{fig:condensatesQM}
\end{figure}

See \cref{fig:condensatesQM} for a comparison of the scaling in the small and large SUSY-breaking regimes in terms of both $\Lambda_\text{SUSY}$ and $\Lambda_\text{QCD}$.

\subsection{S-confinement Phase }
The properly normalized squark condensate is
\begin{align}
    \expval{\tilde{q}^*\tilde{q}} =& \Lambda^2 \left(\frac{N_f-3}{N_f-1} \frac{ m}{\lambda \Lambda}\right)^{\frac{1}{N_f-2}},
\end{align}
The quark condensate is given by
\begin{align}
    \expval{\bar q q} =& -m\Lambda^2 
    \frac{1+(N_f-3)^{\frac{N_f-1}{N_f-2}}}{(N_f-1)^{\frac{1}{N_f-2}}}
    \left(\frac{m}{\lambda \Lambda}\right)^{\frac{1}{N_f-2}}.
\end{align}
Note that the scaling behavior of the condensates differs by a factor of $m$ in agreement with expectations:
\begin{align}
    \expval{\tilde{q}^* q} \propto&\ \Lambda^2\left(\frac{m}{\Lambda}\right)^{\frac{1}{N_f-2}},\label{eq:s-confinesquarkcondense}\\
    \expval{\bar q q}\propto&\ m\Lambda^2\left(\frac{m}{\Lambda}\right)^{\frac{1}{N_f-2}}.\label{eq:s-confinequarkcondense}
\end{align}
For the gluon and gluino condensates, we get:
\begin{align}
    \expval{GG} =& 
    -8
    \frac{(N_f-3)^2(2N_f-1)}{N_f-1}
    \left(
    \frac{N_f-3}{N_f-1}
    \lambda^{-\frac{2}{N_f}}
    \frac{m}{\Lambda}
    \right)^{\frac{N_f}{N_f-2}}
    m\Lambda^3 \ln\frac{\Lambda}{\mu}\ ,
    \\
    \expval{\lambda\lambda} =&  8 \frac{N_f-3}{N_f-2}
    \left(
    \frac{N_f-3}{N_f-1} \lambda^{-\frac{2}{N_f}}
    \frac{m}{\Lambda}
    \right)^{\frac{N_f}{N_f-2}}
    \Lambda^3 \ln\frac{\Lambda}{\mu}\ .
\end{align}
Note that, again, the condensates differ in scaling by a factor of $m$:
\begin{align}
    \expval{\lambda\lambda}\propto&\ m^{\frac{N_f}{N_f-2}}\Lambda^{3-\frac{N_f}{N_f-2}},\label{eq:s-confinegluinocondense}\\
    \expval{GG}\propto&\ m^{1+\frac{N_f}{N_f-2}}\Lambda^{3-\frac{N_f}{N_f-2}},\label{eq:s-confinegluoncondense}
\end{align}
conforming with expectations.

\begin{figure}
    \centering
    \includegraphics[width=\linewidth]{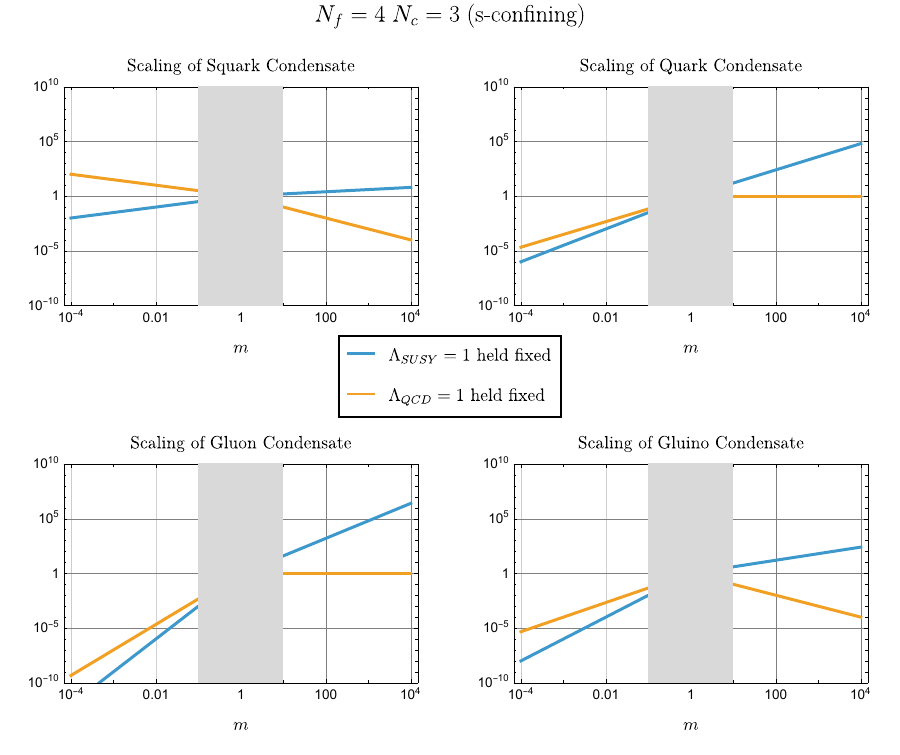}
    \caption{The scaling behavior of the various bilinear condensates in the AMSB-perturbed s-confining phase is shown as a power law in $m$ based on~\cref{eq:s-confinesquarkcondense,eq:s-confinequarkcondense,eq:s-confinegluoncondense,eq:s-confinegluinocondense} in units with $\Lambda_{SUSY}=1$ held fixed (blue) and $\Lambda_{QCD}=1$ held fixed (orange). The latter corresponds to exchanging $\Lambda_{SUSY}$ for $\Lambda_{QCD}$ via \cref{eq:scalematch} which for this particular plot is $\Lambda_{SUSY} = (\Lambda_{QCD}^5/m^2)^{1/3}$.  The choice $N_f=4,N_c=3$ serves as a representative example; other values do not qualitatively alter the plots (so long as $N_f=N_c+1$). The gray region around $m\sim 1$ indicates the incalculable regime. The results in the $m\ll \Lambda$ region are \textit{exact} (up to unknown $\mathcal{O}(1)$ factors not relevant for the scaling behavior) coming from the AMSB calculation. The results in the $m\gg\Lambda$ region are, up to an unknown overall factor, uniquely determined by dimensional analysis (QCD only has one scale), scale matching as in \cref{eq:scalematch}, and the two-loop generation of superpartner condensates in \cref{fig:factorm} (which should be very accurate in the $m\gg\Lambda$ limit, and only inherits the uncertainty in the overall factor of the QCD condensates).}
    \label{fig:condensatesS}
\end{figure}

See \cref{fig:condensatesS} for a comparison of the scaling in the small and large SUSY-breaking regimes in terms of both $\Lambda_\text{SUSY}$ and $\Lambda_\text{QCD}$.

\subsection{Free Magnetic Phase }
The properly normalized squark condensate is
\begin{align}\label{eq:Freemagsquarkcondense}
    \expval{\tilde{q}^*\tilde{q}} =& \frac{\Lambda^2}{\kappa}\left(
    \frac{3N_c-2N_f}{N_c}\frac{m}{\Lambda}
    \right)^{\frac{N_c}{2N_c-N_f}-1} .
\end{align}
We obtain for the quark condensate
\begin{align}\label{eq:Freemagquarkcondense}
    \expval{\bar q q} =& -\frac{\Lambda^3}{\kappa}\left(\frac{N_c}{3N_c-2N_f}-\kappa^2\right)\left(\frac{3N_c-2N_f}{N_c}\frac{m}{\Lambda}\right)^{\frac{N_c}{2N_c-N_f}} .
\end{align}
Note that they differ in scaling by a factor of $m$ in agreement with expectations:
\begin{align}
    \expval{\tilde{q}^*\tilde{q}} \propto&\ m^{\frac{N_c}{2N_c-N_f}-1}\Lambda^{3-\frac{N_c}{2N_c-N_f}}, \label{eq:Freemagsquarkcondense}\\
    \expval{\bar q q}\propto&\ m^{\frac{N_c}{2N_c-N_f}}\Lambda^{3-\frac{N_c}{2N_c-N_f}} \label{eq:Freemagquarkcondense}.
\end{align}
For the gluon and gluino condensates we obtain
\begin{align}
    \expval{GG}=&
    -8\frac{3N_c-2N_f}{2N_c-N_f}(2N_c-N_f\kappa^2)
    \left(
    \frac{3N_c-2N_f}{N_c}
    \frac{m}{\Lambda}
    \right)^{\frac{2N_c}{2N_c-N_f}}
    \Lambda^4 \ln\frac{\Lambda}{\mu}\ \label{eq:Freemaggluoncondense},
    \\
    \expval{\lambda\lambda} =& 
    -8\frac{N_c(N_f-N_c)}{2N_c-N_f}
    \left(
    \frac{3N_c-2N_f}{N_c}
    \frac{m}{\Lambda}
    \right)^{\frac{2N_c}{2N_c-N_f}}\frac{\Lambda^4}{m}\ln\frac{\Lambda}{\mu}\ \label{eq:Freemaggluinocondense}.
\end{align}
Note that the condensates differ by a factor of $m$ in scaling:
\begin{align}
    \expval{\lambda\lambda}\propto&\ m^{\frac{N_f}{2N_c-N_f}}\Lambda^{4-\frac{2N_c}{2N_c-N_f}}, \label{eq:Freemaggluinocondense}\\
    \expval{GG}\propto&\ m^{1+\frac{N_f}{2N_c-N_f}}\Lambda^{4-\frac{2N_c}{2N_c-N_f}}, \label{eq:Freemaggluoncondense}
\end{align}
conforming with expectations.

\begin{figure}
    \centering
    \includegraphics[width=\linewidth]{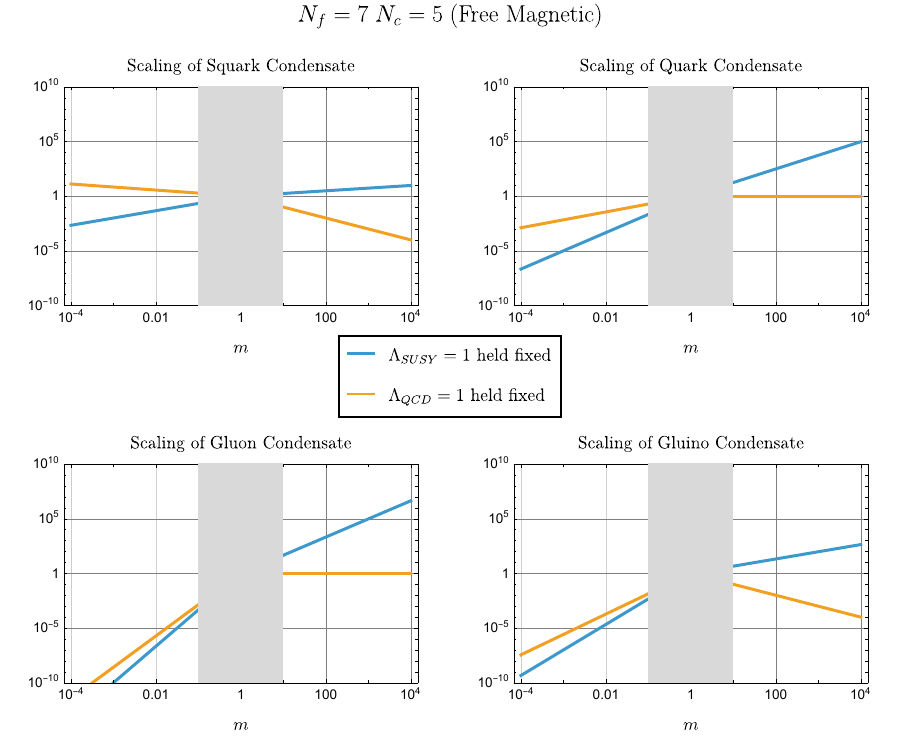}
    \caption{Scaling of the various bilinear condensates in the AMSB-perturbed Free-Magnetic phase, as a power law in $m$ based on~\cref{eq:Freemagsquarkcondense,eq:Freemagquarkcondense,eq:Freemaggluoncondense,eq:Freemaggluinocondense} in units with $\Lambda_{SUSY}=1$ held fixed (blue) and $\Lambda_{QCD}=1$ held fixed (orange). The latter corresponds to exchanging $\Lambda_{SUSY}$ for $\Lambda_{QCD}$ via \cref{eq:scalematch} which for this particular plot is $\Lambda_{SUSY} = (\Lambda_{QCD}^{41}/m^{17})^{1/24}$.  The choice $N_f=7,N_c=5$ is just a representative example -- other values do not qualitatively alter the plots (so long as $N_c+2 \leq N_f \leq 3N_c/2$). The gray region around $m\sim 1$ reflects the incalculability in that region. The results in the $m\ll \Lambda$ region are \textit{exact} (up to unknown $\mathcal{O}(1)$ factors not relevant for the scaling behavior) coming from the AMSB calculation. The results in the $m\gg\Lambda$ region are, up to an unknown overall factor, uniquely determined by dimensional analysis (QCD only has one scale), scale matching as in \cref{eq:scalematch}, and the two-loop generation of superpartner condensates in \cref{fig:factorm} (which should be very accurate in the $m\gg\Lambda$ limit, and only inherits the uncertainty in the overall factor of the QCD condensates).}
    \label{fig:condensatesFM}
\end{figure}

See \cref{fig:condensatesFM} for a comparison of the scaling in the small and large SUSY-breaking regimes in terms of both $\Lambda_\text{SUSY}$ and $\Lambda_\text{QCD}$.

\subsection{Conformal Window}
To understand the impact of AMSB which is expected to deflect the flow away from the IR fixed point, we need to understand how the couplings approach the IR fixed point from their UV values. For the bulk of the conformal window, we lack analytical methods to determine this behavior. Therefore, we study the lower and upper edges of the conformal window and interpolate the behavior in between.

\subsubsection{Lower Conformal Window}
Near the lower edge of the conformal window where $\tilde{\epsilon}\ll 1$, we find 
\begin{align}
    \expval{\tilde{q}^*\tilde{q}}\simeq& \left[\frac{\lambda^{2\frac{N_f}{\tilde{N}_c}} \alpha(2+\alpha)}{4 N_f^{3-\frac{N_f}{\tilde{N}_c}}}\frac{m^2}{\Lambda^2}\right]^{\frac{1}{2-\alpha}} \Lambda^2,\label{eq:LCWsquarkcondense}
\end{align}
\begin{align}
    \expval{\bar{q} q} \simeq&  \left[
    \frac{\lambda^{2\frac{N_f}{\tilde{N}_c}}   \alpha(2+\alpha)}{16 N_f^{3-\frac{N_f}{\tilde{N}_c}}}
    \frac{m^2}{\Lambda^2}\right]^{\frac{2}{2-\alpha}} \Lambda^3 .\label{eq:LCWquarkcondense}
    \end{align}
For the gluon and gluino condensates we obtain
\begin{align}
    \expval{GG}\simeq& -16\frac{\alpha}{2-\alpha}\left[\frac{\lambda^{N_f/\tilde{N}_c} \alpha(\alpha+2)}{4 N_f^{\frac{2\tilde{N}_c-N_f}{2\tilde{N}_c}}}\frac{m^2}{\Lambda^2}\right]^{\frac{4}{2-\alpha}} \Lambda^4 \ln\frac{\Lambda}{\mu}\ ,\label{eq:LCWgluoncondense}
\end{align}
\begin{align}
    \expval{\lambda\lambda}\simeq& -4 \left(
    3N_c - \frac{4-3\alpha}{2-\alpha}N_f^{\frac{2}{2-\alpha}}
    \right) \left[
    \frac{\alpha(2+\alpha)    ^{\frac{N_f}{N_f-N_c}}}{4} (\lambda^2 N_f)^{\frac{2N_c-N_f}{N_f-N_c}}
    \frac{m^2}{\Lambda^2}
    \right]^{\frac{N_f}{(N_f-N_c)(2-\alpha)}} \Lambda^3 \ln\frac{\Lambda}{\mu}\ .\label{eq:LCWgluinocondense}
\end{align}
Note that the scaling relationship holds to leading order in perturbative expansion, but not at sub-leading order.

\begin{figure}
    \centering
    \includegraphics[width= \linewidth]{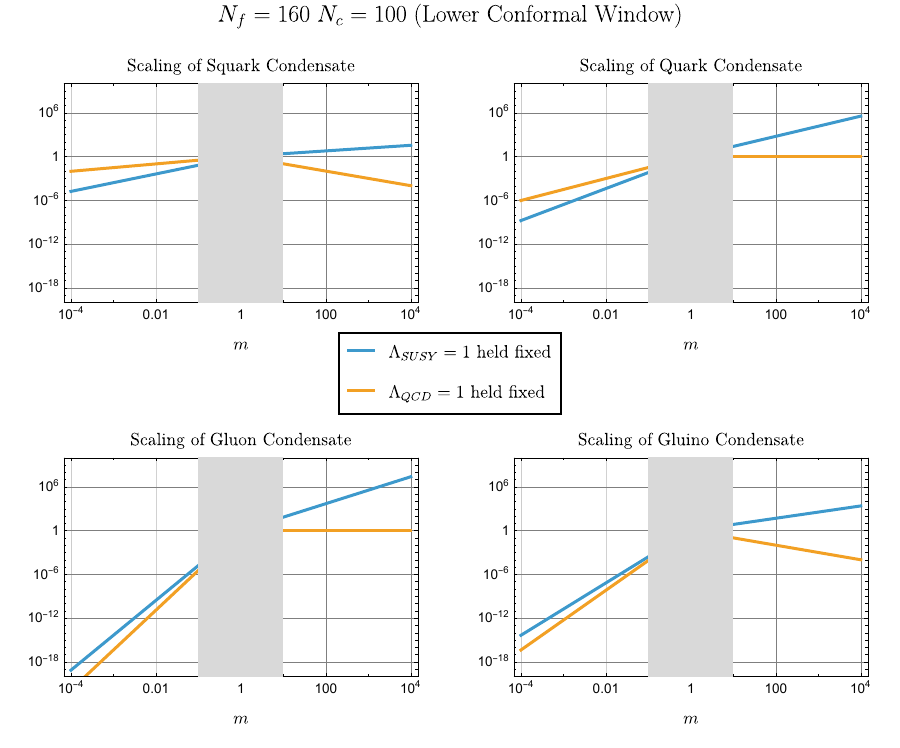}
    \caption{Scaling of the different bilinear condensates in the AMSB-perturbed Lower Conformal Window, as a power law in $m$ based on~\cref{eq:LCWsquarkcondense,eq:LCWquarkcondense,eq:LCWgluoncondense,eq:LCWgluinocondense} in units with $\Lambda_{SUSY}=1$ held fixed (blue) and $\Lambda_{QCD}=1$ held fixed (orange). The latter corresponds to exchanging $\Lambda_{SUSY}$ for $\Lambda_{QCD}$ via \cref{eq:scalematch} which for this particular plot is $\Lambda_{SUSY}=(\Lambda_{QCD}^{13}/m^6)^{1/7}$.  The choice $N_f=151,N_c=100$ is just a representative example -- other values do not qualitatively alter the plots (so long as $3N_c/2 < N_f \ll 3N_c$), but large $N_c$ and small $\tilde{\epsilon}$ are required for the validity of the expansion around the Banks-Zaks fixed point. The gray region around $m\sim 1$ reflects the incalculability in that region. Note that the vertical axis in these plots is changed relative to previous phases to better illustrate the small scaling behavior at low $m$. The results in the $m\ll \Lambda$ region are \textit{exact} (up to unknown $\mathcal{O}(1)$ factors not relevant for the scaling behavior) coming from the AMSB calculation. The results in the $m\gg\Lambda$ region are, up to an unknown overall factor, uniquely determined by dimensional analysis (QCD only has one scale), scale matching as in \cref{eq:scalematch}, and the two-loop generation of superpartner condensates in \cref{fig:factorm} (which should be very accurate in the $m\gg\Lambda$ limit, and only inherits the uncertainty in the overall factor of the QCD condensates).}
    \label{fig:condensatesLCW}
\end{figure}

See \cref{fig:condensatesLCW} for a comparison of the scaling in the small and large SUSY-breaking regimes in terms of both $\Lambda_\text{SUSY}$ and $\Lambda_\text{QCD}$.

\subsubsection{Upper Conformal Window}
Near the upper edge of the conformal window, where $\epsilon \ll 1$, we find
\begin{align}
    \expval{\tilde{q}^*\tilde{q}} =& \sqrt{\frac{1+\epsilon}{3N_c}}\Lambda^2 \left(\frac{m}{Y\Lambda}\right)^{\frac{1}{\epsilon}},\label{eq:UCWsquarkcondense}
    \\
    \expval{\bar{q}q} =& \frac{\sqrt{1+\epsilon}Y}{\sqrt{3N_c   }}\Lambda^3\left(
    \frac{m}{Y\Lambda}
    \right)^{1+\frac{1}{\epsilon}}.\label{eq:UCWquarkcondense}
\end{align}
In the limit that $\epsilon \ll 1$, such that $\alpha\sim 21\epsilon^2 \ll \epsilon$, we find to leading order
\begin{align}
    \expval{GG} \simeq&\, 16\frac{1-\epsilon}{\epsilon } \left(\frac{m}{Y \Lambda}\right)^{2+\frac{2}{\epsilon}} \Lambda^4 \ln\frac{\Lambda}{\mu}\ ,\label{eq:UCWgluoncondense}
    \\
    \expval{\lambda\lambda} \simeq&\, 8\frac{1-\epsilon}{\epsilon} \left( \frac{ m}{Y\Lambda}\right)^{1+\frac{2}{\epsilon}} \Lambda^3 \ln\frac{\Lambda}{\mu}\ .\label{eq:UCWgluinocondense}
\end{align}
Note that this is not necessarily the leading order term when $\epsilon$ and $m/\Lambda$ are comparable. Different such regimes can have results that differ by powers sub-leading in $\epsilon$.

\begin{figure} 
    \centering
    \includegraphics[width=\linewidth]{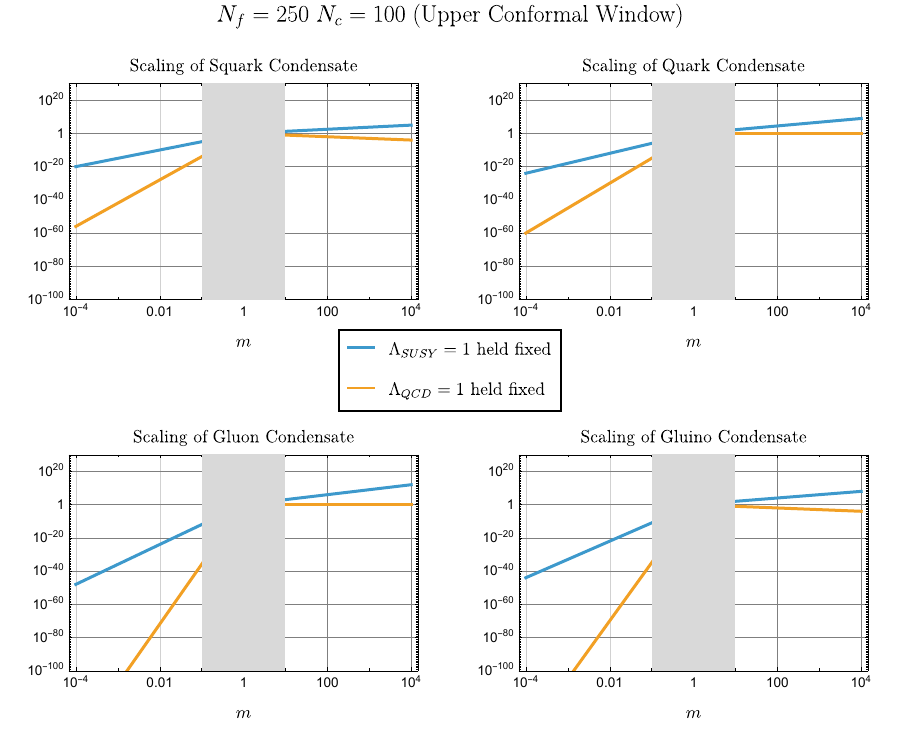}
    \caption{Scaling of the various bilinear condensates in the AMSB-perturbed Lower Conformal Window, as a power law in $m$ based on~\cref{eq:UCWsquarkcondense,eq:UCWquarkcondense,eq:UCWgluoncondense,eq:UCWgluinocondense} in units with $\Lambda_{SUSY}=1$ held fixed (blue) and $\Lambda_{QCD}=1$ held fixed (orange). The latter corresponds to exchanging $\Lambda_{SUSY}$ for $\Lambda_{QCD}$ via \cref{eq:scalematch} which for this particular plot is $\Lambda_{SUSY}=\Lambda_{QCD}^4/m^3$.  The choice $N_f=250,N_c=100$ is just a representative example -- other values do not qualitatively alter the plots (so long as $3N_c/2 \ll N_f < 3N_c$), but large $N_c$ and small ${\epsilon}$ are required for the validity of the expansion around the Banks-Zaks fixed point. The gray region around $m\sim 1$ reflects the incalculability in that region. Note that the vertical axis is expanded by many orders of magnitude compared to the previous phases, in order to show the extremely small scaling behavior at low $m$. The results in the $m\ll \Lambda$ region are \textit{exact} (up to unknown $\mathcal{O}(1)$ factors not relevant for the scaling behavior) coming from the AMSB calculation. The results in the $m\gg\Lambda$ region are, up to an unknown overall factor, uniquely determined by dimensional analysis (QCD only has one scale), scale matching as in \cref{eq:scalematch}, and the two-loop generation of superpartner condensates in \cref{fig:factorm} (which should be very accurate in the $m\gg\Lambda$ limit, and only inherits the uncertainty in the overall factor of the QCD condensates).}
    \label{fig:condensatesUCW}
\end{figure}

See \cref{fig:condensatesUCW} for a comparison of the scaling in the small and large SUSY-breaking regimes in terms of both $\Lambda_\text{SUSY}$ and $\Lambda_\text{QCD}$.

\FloatBarrier
\section{Calculation of Mass Spectrum}\label{sec:massspectrum}

In this section, we present the mass spectrum of low-lying states. We warn the readers that the crossover does not necessarily guarantee that the low-lying spectra continuously change from the near-SUSY to the non-SUSY limit except for the massless states (Nambu--Goldstone bosons). Yet we observe that the $0^+$ states of the meson superfield appear to show the same feature in both limits. The $0^+$ states are established experimentally only rather recently, and it is curious that we can understand their spectrum.

We show the result based on the specific $N_c=3$ case. We hope that we can extract lessons from these results. The flow chart of the calculation of this section is in the following way. From the calculation of the potential from superpotential in \cref{sec:vacua}, we know the vacuum and its field values. We introduce fluctuation of the fields around the vacuum, and then we calculate the mass matrix by taking second derivatives with respect to the fields. We decomposed the meson field $M = M^a T^a$ into real part and imaginary part 
        $M^a=\frac{1}{\sqrt{2}}(\sigma^a+i\pi^a).$ We turn on the finite quark mass to see how it affects, we treated the quark mass as a perturbation much smaller than the AMSB scale $m$. We consider the region such that the vacuum is slightly perturbed by the quark mass to avoid run-away direction. We denote the configuration of the minimum by $\sigma_0+\delta\sigma$. We consider the potential as vacuum (massless) part and finite (quark) mass part $V=V_0+V_{\text{fin}}$. In order for this configuration to be realized as a vacuum, the potential should satisfy the stationary condition
        \begin{align}
            0&=\frac{\partial V}{\partial M}|_{M=\sigma_0+\delta\sigma}=\frac{\partial V_0}{\partial M}|_{M=\sigma_0+\delta\sigma}+\frac{\partial V_{\text{fin}}}{\partial M}|_{M=\sigma_0+\delta\sigma}
            \simeq
            \frac{\partial^2 V_0}{\partial M^2}(\sigma_0)\delta\sigma+\frac{\partial V_{\text{fin}}}{\partial M}(\sigma_0),\\
            \delta\sigma&\simeq -(\text{Mass}^2)^{-1}\frac{\partial V_{\text{fin}}}{\partial M}.
        \end{align}
Then we can evaluate the mass at the point $M=\sigma_0+\delta\sigma$. We note that the second derivative of the potential is nothing but the mass square matrix.

We have derived expressions for the scalar and pseudoscalar masses \textit{without} quark masses, in the ADS, Quantum-Modified, and s-confining phases, for arbitrary $N_f,N_c$. For the ADS case we find:
\begin{align}
    \text{Singlet scalar}: \frac{2N_c(3N_c-N_f)^2}{(N_c-N_f)^2(N_c+N_f)}m^2,\ 
    \text{Adjoint scalar}: \frac{2N_c(3N_c-N_f)^2}{(N_c+N_f)^2}m^2,\nonumber\\
    \text{Singlet pseudo-scalar}: \frac{2N_f(3N_c-N_f)^2}{(N_c-N_f)^2(N_c+N_f)}m^2,\ 
    \text{Adjoint pseudo-scalar}: 0. \label{eq:ADSvacuumMass}
\end{align}
For the Quantum-Modified case we find:
\begin{align}
    \text{Singlet scalar}&: N\lambda^2\Lambda^2 \nonumber\\
    \text{Adjoint scalar}&: 2m^2\nonumber\\
    \text{Singlet pseudo-scalar}&: N\lambda^2\Lambda^2\nonumber\\
    \text{Adjoint pseudo-scalar}&: 0. \label{eq:QMvacuumMass}
\end{align}
And for the s-confining case, we find:
\begin{align}
    \text{Singlet scalar}&: \frac{N_f^2(N_f-3)^2}{4(N_f-1)}m^2,\ 
    \text{Adjoint scalar}: \frac{2(N_f-3)^2}{(N_f-1)^2}m^2\nonumber\\
    \text{Singlet pseudo-scalar}&: \frac{N_f(N_f-2)(N_f-3)^2}{4(N_f-1)}m^2,\ \text{Adjoint pseudo-scalar}: 0. \label{eq:SvacuumMass}
\end{align}
Below, we calculate the effect of finite quark masses in the specific cases of $(N_f,N_c) = (3,4),\, (3,3),$ and $(4,3)$. We find mass spectra that satisfy a sum rule
\begin{align}
    m_\sigma^2 + c\, m_\pi^2 = m_0^2,
    \label{eq:sumrule}
\end{align}
where $m_0^2$ and $c$ are model-dependent constants common to the entire adjoint scalars and pseudo-scalars, which then we can compare to the real QCD data.

\subsection{Case Study: ADS Case $N_f=3$ and $N_c=4$}\label{sec:Nf3Nc4}
We have the superpotential as in \cref{eq:WADS}, which we copy here,
\begin{align}
    W&= \frac{\Lambda^9}{\det M}-\operatorname{Tr}[m_QM].
\end{align}
For the massless case ($m_Q=0$), we obtained the following mass spectrum
\begin{align}
    \text{Singlet scalar}: \frac{648}{7}m^2,\ 
    \text{Octet scalar}: \frac{162}{49}m^2,\\
    \text{Singlet pseudo-scalar}: \frac{486}{7}m^2,\ 
    \text{Octet pseudo-scalar}: 0.
\end{align}
consistent with the general formulae \cref{eq:ADSvacuumMass}. When we turn on a finite quark mass, we can obtain the following mass matrix for $\sigma$
\begin{align}
    m_{\sigma}^2&=\begin{pmatrix}
        m^2_{\sigma 1}&0&0&0&0&0&0&0&0\\
        0&m_{\sigma 1}^2&0&0&0&0&0&0&0\\
        0&0&m_{\sigma 1}^2&0&0&0&0&m_{\sigma38}^2&m_{\sigma03}^2\\
        0&0&0&m_{\sigma 2}^2&0&0&0&0&0\\
        0&0&0&0&m_{\sigma 2}^2&0&0&0&0\\
        0&0&0&0&0&m^2_{\sigma 3}&0&0&0\\
        0&0&0&0&0&0&m_{\sigma 3}^2&0&0\\
        0&0&m^2_{\sigma38}&0&0&0&0&m_{\sigma 8}^2&m_{\sigma08}^2\\
        0&0&m^2_{\sigma03}&0&0&0&0&m^2_{\sigma08}&m_{\sigma0}^2\\
    \end{pmatrix},
\end{align}
where
\begin{align}
    m^2_{\sigma 1}&=\frac{162}{49}m^2-\frac{19}{49}m\left(m_{Q1}+m_{Q2}-\frac{128}{19}m_{Q3}\right),\nonumber\\
    m^2_{\sigma 2}&=\frac{162}{49}m^2-\frac{19}{49}m\left(-\frac{128}{19}m_{Q1}+m_{Q2}+m_{Q3}\right),\nonumber\\
    m^2_{\sigma 3}&=\frac{162}{49}m^2-\frac{19}{49}m\left(m_{Q1}-\frac{128}{19}m_{Q2}+m_{Q3}\right),\nonumber\\
    m^2_{\sigma 8}&= \frac{162}{49}m^2+\frac{79}{49}m\left(m_{Q1}+m_{Q2}-\frac{68}{79}m_{Q3}\right),\nonumber\\
    m^2_{\sigma 0}&= \frac{648}{7}m^2+\frac{232}{7}m(m_{Q1}+m_{Q2}+m_{Q3}),\nonumber\\
    m^2_{\sigma38}&=-\sqrt{3}m(m_{Q1}-m_{Q2}),\nonumber\\
    m^2_{\sigma03}&=\frac{15\sqrt{6}}{2}m(m_{Q1}-m_{Q2})\nonumber,\\
    m^2_{\sigma08}&=\frac{15\sqrt{2}}{2}m(m_{Q1}+m_{Q2}-2m_{Q3}).\nonumber
\end{align}
We can obtain the eigenvalues of this mass squared matrix. We write the eigenvalues subtracted by a common piece $49m(m_{Q1}+m_{Q2}+m_{Q3})/128$ as
\begin{align}
    \text{singlet}&: \frac{648}{7}m^2+\frac{1496}{49}m(m_{Q1}+m_{Q2}+m_{Q3}),\nonumber\\
    \sigma_{3,8}&: \frac{162}{49}m^2-2m(m_{Q1}+m_{Q2}+m_{Q3}),\nonumber\\
   \sigma_{1,2}&: \frac{162}{49}m^2-3m(m_{Q1}+m_{Q2}),\nonumber\\
   \sigma_{4,5}&: \frac{162}{49}m^2-3m(m_{Q1}+m_{Q3}),\nonumber\\
   \sigma_{6,7}&: \frac{162}{49}m^2-3m(m_{Q2}+m_{Q3}).
\end{align}
The mass-square matrix for $\pi$ is
\begin{align}
    m_{\pi}^2&=\begin{pmatrix}
        m^2_{\pi 1}&0&0&0&0&0&0&0&0\\
        0&m_{\pi 1}^2&0&0&0&0&0&0&0\\
        0&0&m_{\pi 1}^2&0&0&0&0&m_{\pi 38}^2&m_{\pi 03}^2\\
        0&0&0&m_{\pi 2}^2&0&0&0&0&0\\
        0&0&0&0&m_{\pi 2}^2&0&0&0&0\\
        0&0&0&0&0&m^2_{\pi 3}&0&0&0\\
        0&0&0&0&0&0&m_{\pi 3}^2&0&0\\
        0&0&m^2_{\pi38}&0&0&0&0&m_{\pi 8}^2&m_{\pi08}^2\\
        0&0&m^2_{\pi03}&0&0&0&0&m^2_{\pi08}&m_{\pi0}^2\\
    \end{pmatrix},
\end{align}
where
\begin{align}
    m^2_{\pi 1}&=\frac{25}{7}m(m_{Q1}+m_{Q2}),\nonumber\\
    m^2_{\pi 2}&=\frac{25}{7}m(m_{Q2}+m_{Q3}),\nonumber\\
    m^2_{\pi 3}&=\frac{25}{7}m(m_{Q3}+m_{Q1}),\nonumber\\
    m^2_{\pi 8}&= \frac{25}{21}m(m_{Q1}+m_{Q2}+4m_{Q3}),\nonumber\\
    m^2_{\pi 0}&= \frac{486}{7}m^2+\frac{872}{21}(m_{Q1}+m_{Q2}+m_{Q3}),\nonumber\\
    m^2_{\pi38}&=\frac{25\sqrt{3}}{21}m(m_{Q1}-m_{Q2}),\nonumber\\
    m^2_{\pi03}&=\frac{11\sqrt{6}}{42}m(m_{Q1}-m_{Q2}),\nonumber\\
    m^2_{\pi08}&=\frac{11\sqrt{2}}{42}m(m_{Q1}+m_{Q2}-2m_{Q3}).\nonumber
\end{align}
The mass eigenvalues are 
\begin{align}
    \text{singlet}&: \frac{486}{7}m^2+\frac{872}{21}m(m_{Q1}+m_{Q2}+m_{Q3}),\nonumber\\
    \pi_{3,8}&: \frac{50}{21}m(m_{Q1}+m_{Q2}+m_{Q3},)\nonumber\\
   \pi_{1,2}&: \frac{25}{7}m(m_{Q1}+m_{Q2}),\nonumber\\
   \pi_{4,5}&: \frac{25}{7}m(m_{Q2}+m_{Q3}),\nonumber\\
   \pi_{6,7}&: \frac{25}{7}m(m_{Q3}+m_{Q1}).
\end{align}
From these spectra, we can see the sum rule \cref{eq:sumrule} for the whole octet by working out
\begin{align}
m^2_\sigma+\frac{21}{25} m^2_\pi=
    \frac{162}{49}m^2
    -\frac{49}{128}m(m_{Q1}+m_{Q2}+m_{Q3}). \nonumber
    \label{eq:sumrule1}
\end{align}
We have also checked that the supertrace of the spectrum vanishes.

\subsection{Case Study: Quantum Modified Case $N_f=3$ and $N_c=3$}
We have the superpotential as in \cref{eq:WQM}. For this analysis, we assume canonical and minimal K\"ahler potential. We write the expression again here
\begin{align}
W&=\lambda X\left(\frac{\det M}{\Lambda^{N_f-2}}-\Lambda^2\right)    
\end{align}
In the massless case ($m_Q=0$), we obtained the following mass spectrum
\begin{align}
    \text{Singlet scalar}&: 3\lambda^2\Lambda^2 \nonumber\\
    \text{Octet scalar}&: 2m^2\nonumber\\
    \text{Singlet pseudo-scalar}&: 3\lambda^2\Lambda^2\nonumber\\
    \text{Octet pseudo-scalar}&: 0.
\end{align}
This result is consistent with \cref{eq:QMvacuumMass}. When we turn on a finite quark mass, we can obtain the following mass matrix for the scalar $\sigma$
\begin{align}
m_{\sigma}^2&=
\begin{pmatrix}
m_{X}^2&0&0&m_{X3}^2&0&0&0&0&m_{X8}^2&m_{X0}^2\\
0&m^2_{\sigma1}&0&0&0&0&0&0&0&0\\
0&0&m_{\sigma1}^2&0&0&0&0&0&0&0\\
m_{X3}^2&0&0&m_{\sigma1}^2&0&0&0&0&m_{\sigma38}^2&m_{\sigma03}^2\\
0&0&0&0&m_{\sigma2}^2&0&0&0&0&0\\
0&0&0&0&0&m_{\sigma2}^2&0&0&0&0\\
0&0&0&0&0&0&m^2_{\sigma3}&0&0&0\\
0&0&0&0&0&0&0&m_{\sigma3}^2&0&0\\
m^2_{X8}&0&0&m^2_{\sigma38}&0&0&0&0&m_{\sigma8}^2&m_{\sigma08}^2\\
m^2_{X0}&0&0&m^2_{\sigma03}&0&0&0&0&m^2_{\sigma08}&m_{\sigma0}^2\\
\end{pmatrix},
\end{align}
where
\begin{align}
    m^2_{X}&=3\lambda^2\Lambda^2-4m^2+\frac{4}{3}m(m_{Q1}+m_{Q2}+m_{Q3}),\nonumber\\
m_{X3}^2&=\frac{9\lambda^2\Lambda^2-5m^2}{6\sqrt{2}\lambda\Lambda}(m_{Q1}-m_{Q2}),\nonumber\\
m_{X8}^2&=\frac{9\lambda^2\Lambda^2-5m^2}{6\sqrt{6}\lambda\Lambda}(m_{Q1}+m_{Q2}-2m_{Q3}),\nonumber\\
m^2_{X0}&=3\sqrt{3}\lambda m\Lambda +\frac{6\lambda^2\Lambda^2-4m^2}{3\sqrt{3}\lambda\Lambda}(m_{Q1}+m_{Q2}+m_{Q3}),\nonumber\\
    m^2_{\sigma1}&=2m^2-\frac{5}{6}m\left(m_{Q1}+m_{Q2}-\frac{16}{5}m_{Q3}\right),\nonumber\\
    m^2_{\sigma 2}&=2m^2-\frac{5}{6}m\left(-\frac{16}{5}m_{Q1}+m_{Q2}+m_{Q3}\right),\nonumber\\
    m^2_{\sigma 3}&=2m^2-\frac{5}{6}m\left(m_{Q1}-\frac{16}{5}m_{Q2}+m_{Q3}\right),\nonumber\\
    m^2_{\sigma 8}&= 2m^2+\frac{3}{2}m\left(m_{Q1}+m_{Q2}-\frac{4}{3}m_{Q3}\right),\nonumber\\
    m^2_{\sigma 0}&= 3\lambda^2\Lambda^2-2m^2+\frac{14}{3}m(m_{Q1}+m_{Q2}+m_{Q3}),\nonumber\\
    m^2_{\sigma38}&=-\frac{7}{2\sqrt{3}}m(m_{Q1}-m_{Q2}),\nonumber\\
    m^2_{\sigma03}&=-\frac{\sqrt{3}(\lambda^2\Lambda^2-3m^2)}{2\sqrt{2}m}(m_{Q1}-m_{Q2})\\
    m^2_{\sigma08}&=-\frac{\lambda^2\Lambda^2-3m^2}{2\sqrt{2}m}(m_{Q1}+m_{Q2}-2m_{Q3}).\nonumber
\end{align}

When we turn on a finite quark mass, we can obtain the following mass matrix for the pseudo-scalar $\pi$
\begin{align}
m_{\pi}^2&=
\begin{pmatrix}
m_{X}^2&0&0&m_{X3}^2&0&0&0&0&m_{X8}^2&m_{X0}^2\\
0&m^2_{\pi1}&0&0&0&0&0&0&0&0\\
0&0&m_{\pi1}^2&0&0&0&0&0&0&0\\
m_{X3}^2&0&0&m_{\pi1}^2&0&0&0&0&m_{\pi38}^2&m_{\pi03}^2\\
0&0&0&0&m_{\pi2}^2&0&0&0&0&0\\
0&0&0&0&0&m_{\pi2}^2&0&0&0&0\\
0&0&0&0&0&0&m^2_{\pi3}&0&0&0\\
0&0&0&0&0&0&0&m_{\pi3}^2&0&0\\
m^2_{X8}&0&0&m^2_{\pi38}&0&0&0&0&m_{\pi8}^2&m_{\pi08}^2\\
m^2_{X0}&0&0&m^2_{\pi03}&0&0&0&0&m^2_{\pi08}&m_{\pi0}^2\\
\end{pmatrix},
\end{align}
where
\begin{align}
    m^2_{X}&=3\lambda^2\Lambda^2-4m^2+\frac{4}{3}m(m_{Q1}+m_{Q2}+m_{Q3}),\nonumber\\
m_{X3}^2&=-\frac{9\lambda^2\Lambda^2-5m^2}{6\sqrt{2}\lambda\Lambda}(m_{Q1}-m_{Q2}),\nonumber\\
m_{X8}^2&=-\frac{9\lambda^2\Lambda^2-5m^2}{6\sqrt{6}\lambda\Lambda}(m_{Q1}+m_{Q2}-2m_{Q3}),\nonumber\\
m^2_{X0}&=3\sqrt{3}\lambda m\Lambda +\frac{2\lambda\Lambda}{\sqrt{3}}(m_{Q1}+m_{Q2}+m_{Q3}),\nonumber\\
    m^2_{\pi1}&=\frac{3}{2}m\left(m_{Q1}+m_{Q2}\right),\nonumber\\
    m^2_{\pi 2}&=\frac{3}{2}m\left(m_{Q2}+m_{Q3}\right),\nonumber\\
    m^2_{\pi 3}&=\frac{3}{2}m\left(m_{Q3}+m_{Q1}\right),\nonumber\\
    m^2_{\pi 8}&=\frac{1}{2}m\left(m_{Q1}+m_{Q2}+4m_{Q3}\right),\nonumber\\
    m^2_{\pi 0}&= 3\lambda^2\Lambda^2+2m^2+\frac{10}{3}m(m_{Q1}+m_{Q2}+m_{Q3}),\nonumber\\
    m^2_{\pi38}&=\frac{\sqrt{3}}{2}m(m_{Q1}-m_{Q2}),\nonumber\\
    m^2_{\pi03}&=-\frac{(3\lambda^2\Lambda^2+m^2)}{2\sqrt{6}m}(m_{Q1}-m_{Q2}),\nonumber\\
    m^2_{\pi08}&=-\frac{3\lambda^2\Lambda^2+m^2}{6\sqrt{2}m}(m_{Q1}+m_{Q2}-2m_{Q3}).\nonumber
\end{align}

We checked that the supertrace vanishes. The fermion mass matrix is the following
\begin{align}
    m_f&=
\begin{pmatrix}
0&0&0&m_{X3}&0&0&0&0&m_{X8}&m_{X0}\\
0&m_{f1}&0&0&0&0&0&0&0&0\\
0&0&m_{f1}&0&0&0&0&0&0&0\\
m_{X3}&0&0&m_{f1}&0&0&0&0&m_{f38}&m_{f03}\\
0&0&0&0&m_{f2}&0&0&0&0&0\\
0&0&0&0&0&m_{f2}&0&0&0&0\\
0&0&0&0&0&0&m_{f3}&0&0&0\\
0&0&0&0&0&0&0&m_{f3}&0&0\\
m_{X8}&0&0&m_{f38}&0&0&0&0&m_{f8}&m_{f08}\\
m_{X0}&0&0&m_{f03}&0&0&0&0&m_{f08}&m_{f0}\\
\end{pmatrix},
\end{align}
where
\begin{align}
m_{X3}&= -\frac{\lambda\Lambda}{2\sqrt{2}m}(m_{Q1}-m_{Q2}),\nonumber\\
m_{X8}&=-\frac{\lambda\Lambda}{2\sqrt{6}m}(m_{Q1}+m_{Q2}-2m_{Q3}),\nonumber\\
m_{X0}&=\sqrt{3}\lambda\Lambda+\frac{2m}{3\sqrt{3}\lambda\Lambda}(m_{Q1}+m_{Q2}+m_{Q3})-\frac{2m^2}{\sqrt{3}\lambda\Lambda},\nonumber\\
m_{f1}&=-m-\frac{1}{6}(m_{Q1}+m_{Q2}+4m_{Q3}),\nonumber\\
m_{f2}&=-m-\frac{1}{6}(4m_{Q1}+m_{Q2}+m_{Q3}),\nonumber\\
m_{f3}&=-m+\frac{1}{6}(m_{Q1}+4m_{Q2}+m_{Q3}),\nonumber\\
m_{f8}&=-m-\frac{1}{2}(m_{Q1}+m_{Q2}),\nonumber\\
m_{f08}&=-\frac{1}{6\sqrt{2}}(m_{Q1}+m_{Q2}-2m_{Q3}),\nonumber\\
m_{f0}&=2m+\frac{2}{3}(m_{Q1}+m_{Q2}+m_{Q3}).
\end{align}
The sum rule \cref{eq:sumrule} holds if we choose $c=7/3$, namely, 
\begin{align}
    m_\sigma^2+\frac{7}{3}m_\pi^2=m_0^2
\end{align}
for the entire octet. We also verified that the supertrace vanishes.

\subsection{Case Study: S-confinement Case $N_f=4$ and $N_c=3$}
We just list up some results. We have the superpotential as in \cref{eq:Wsconfine}
\begin{align}
    W&= \frac{\det M}{\Lambda}-\Lambda \operatorname{Tr}[m_Q M].
\end{align}
For the massless case ($m_Q=0$), we obtained the following mass spectrum
\begin{align}
    \text{Singlet scalar}: \frac{4}{3}m^2,\ 
    \text{Octet scalar}: \frac{2}{9}m^2,\\
    \text{Singlet pseudo-scalar}: \frac{2}{3}m^2,\ 
    \text{Octet pseudo-scalar}: 0.
\end{align}

This result is consistent with \cref{eq:SvacuumMass}. When we turn on a finite quark mass, we can obtain the following mass matrix for $\sigma$. It is $16\times16$ matrix, and huge if we write up in matrix form, so we only write the non-zero components.
\begin{align}
    m^2_{\sigma 11}=m^2_{\sigma22}=m^2_{\sigma33}&=\frac{2}{9}m^2+\frac{\sqrt{3}}{6}\sqrt{m\Lambda}\left(-11m_{Q1}-11m_{Q2}+16m_{Q3}+16m_{Q4}\right),\nonumber\\
m^2_{\sigma44}=m^2_{\sigma55}&=\frac{2}{9}m^2+\frac{\sqrt{3}}{6}\sqrt{m\Lambda}\left(16m_{Q1}-11m_{Q2}-11m_{Q3}+16m_{Q4}\right),\nonumber\\
m^2_{\sigma66}=m^2_{\sigma77}&=\frac{2}{9}m^2+\frac{\sqrt{3}}{6}\sqrt{m\Lambda}\left(-11m_{Q1}+16m_{Q2}-11m_{Q3}+16m_{Q4}\right),\nonumber\\
    m^2_{\sigma88}=m^2_{\sigma1313}=m^2_{\sigma1414}=&= \frac{2}{9}m^2+\frac{\sqrt{3}}{6}\sqrt{m\Lambda}\left(16m_{Q1}+16m_{Q2}-11m_{Q3}-11m_{Q4}\right),\nonumber\\
m^2_{\sigma99}=m^2_{\sigma1010}&= \frac{2}{9}m^2+\frac{\sqrt{3}}{6}\sqrt{m\Lambda}(-11m_{Q1}+16m_{Q2}+16m_{Q3}-11m_{Q4}),\nonumber\\
m^2_{\sigma1111}=m^2_{\sigma1212}&=\frac{2}{9}m^2+\frac{\sqrt{3}}{6}\sqrt{m\Lambda}(16m_{Q1}-11m_{Q2}+16m_{Q3}-11m_{Q4}),\nonumber\\ m^2_{\sigma1515}&=\frac{2}{9}m^2+\frac{5\sqrt{3}}{12}\sqrt{m\Lambda}(m_{Q1}+m_{Q2}+m_{Q3}+m_{Q4}),\nonumber\\
m^2_{\sigma1616}&=\frac{2}{9}m^2+\frac{19\sqrt{3}}{4}\sqrt{m\Lambda}(m_{Q1}+m_{Q2}+m_{Q3}+m_{Q4}),\nonumber\\
m^2_{\sigma315}&=-\frac{9\sqrt{6}}{4}\sqrt{m\Lambda}(m_{Q1}-m_{Q2}),\nonumber\\
m^2_{\sigma316}&=\frac{7\sqrt{6}}{12}\sqrt{m\Lambda}(m_{Q1}-m_{Q2}),\nonumber\\
m^2_{\sigma815}&=-\frac{9\sqrt{6}}{4}\sqrt{m\Lambda}(m_{Q3}-m_{Q4}),\nonumber\\
m^2_{\sigma316}&=\frac{7\sqrt{6}}{12}\sqrt{m\Lambda}(m_{Q3}-m_{Q4}),\nonumber\\
m^2_{\sigma1516}&=\frac{7\sqrt{3}}{12}\sqrt{m\Lambda}(m_{Q1}+m_{Q2}-m_{Q3}-m_{Q4}). \label{eq:sumrule2}
\end{align}

For $\pi$, we obtained the following matrix.
\begin{align}
    m^2_{\pi 11}=m^2_{\pi22}=m^2_{\pi33}&=\frac{7\sqrt{3}}{6}\sqrt{m\Lambda}\left(m_{Q1}-m_{Q2}\right),\nonumber\\
m^2_{\pi44}=m^2_{\pi55}&=\frac{7\sqrt{3}}{6}\sqrt{m\Lambda}\left(m_{Q2}+m_{Q3}\right),\nonumber\\
m^2_{\pi66}=m^2_{\pi77}&=\frac{7\sqrt{3}}{6}\sqrt{m\Lambda}\left(m_{Q1}+m_{Q3}\right),\nonumber\\
m^2_{\pi88}=m^2_{\pi1313}=m^2_{\pi1414}=&= \frac{7\sqrt{3}}{6}\sqrt{m\Lambda}\left(m_{Q3}+m_{Q4}\right),\nonumber\\
m^2_{\pi99}=m^2_{\pi1010}&= \frac{7\sqrt{3}}{6}\sqrt{m\Lambda}(m_{Q1}+m_{Q4}),\nonumber\\
m^2_{\sigma1111}=m^2_{\sigma1212}&=\frac{7\sqrt{3}}{6}\sqrt{m\Lambda}(m_{Q2}+m_{Q4}),\nonumber
\end{align}
\begin{align}
m^2_{\sigma1515}&=\frac{7\sqrt{3}}{12}\sqrt{m\Lambda}(m_{Q1}+m_{Q2}+m_{Q3}+m_{Q4}),\nonumber\\
m^2_{\sigma1616}&=\frac{4}{3}m^2+\frac{17\sqrt{3}}{4}\sqrt{m\Lambda}(m_{Q1}+m_{Q2}+m_{Q3}+m_{Q4}),\nonumber\\
m^2_{\sigma315}&=-\frac{7\sqrt{6}}{12}\sqrt{m\Lambda}(m_{Q1}-m_{Q2}),\nonumber\\
m^2_{\sigma316}&=\frac{9\sqrt{6}}{4}\sqrt{m\Lambda}(m_{Q1}-m_{Q2}),\nonumber\\
m^2_{\sigma815}&=-\frac{7\sqrt{6}}{12}\sqrt{m\Lambda}(m_{Q3}-m_{Q4}),\nonumber\\
m^2_{\sigma316}&=\frac{9\sqrt{6}}{4}\sqrt{m\Lambda}(m_{Q3}-m_{Q4}),\nonumber\\
m^2_{\sigma1516}&=\frac{9\sqrt{3}}{4}\sqrt{m\Lambda}(m_{Q1}+m_{Q2}-m_{Q3}-m_{Q4})
\end{align}
The sum rule \cref{eq:sumrule} holds if we choose $c=27/7$, namely, 
\begin{align}
    m_\sigma^2+\frac{27}{7}m_\pi^2=m_0^2,
\end{align}
for the entire adjoint of $\mathrm{SU}(4)$.

\subsection{Implication of the Sum Rule}
As we have seen from the mass spectrum, the scalar mass and pseudo-scalar mass are summed up in the following form 
\begin{align}
    m_{\sigma}^2=m_0^2-c\,m_\pi^2, \label{eq:sumrule3}
\end{align}
where $c>0$ is a constant, $m_0^2$ is an overall mass scale to be renormalized. We summarize the values of $c$ from \cref{eq:sumrule1,eq:sumrule2,eq:sumrule3} in \cref{tab:masscs}.
\begin{table}[h]
\centering
\begin{tabular}{|c|c|}
\hline
{Models} & {$c$} \\ \hline
{ADS $N_f=3,N_c=4$} & {$21/25$} \\ \hline
{Quantum modified $N_f=3,N_c=3$} & {$7/3$} \\ \hline 
{s-confinement $N_f=4,N_c=3$} & {$27/7$} \\ \hline
\end{tabular}
\caption{The values of the slope $c$ from the sum rule for the explicit cases worked out in the text.}
\label{tab:masscs}
\end{table}


It is curious that the mass spectrum of scalars is upside down compared to that of pseudoscalars because of the sum rule, namely that the isospin triplet is the highest, isodoublet strange states is in the middle, and the isospin singlet is the lowest (see, {\it e.g.}\/, Fig.~8 in \cite{Kondo:2022lgg}). This fact cannot be understood if we regard scalar states as $q\bar{q}$ mesons as an isospin triplet would not contain the strange quark while the isosinglet would. This point was explained by Robert Jaffe \cite{Jaffe:1976ig} that they are actually tetraquark $(q q \bar{q}\bar{q})$ states. Forming $qq$ in an anti-symmetric combination both in color and flavor, the isosinglet combination is $ud$ and hence is light, while the isodoublet combination is $(us,ds)$ and is heavy. Further combining them with $\bar{q}\bar{q}$, one can understand this upside-down spectrum. It was much more recent that lattice calculations advanced to compute the phase shifts in $\pi\pi$ scattering to identify a broad resonance \cite{Briceno:2016mjc,Briceno:2017qmb} using L\"uscher's method using the finite volume effect in the energy of two hadron states \cite{Luscher:1990ux}. In fact the lattice results support the idea that the scalar masses are pushed down as the pseudoscalar masses go up.

\begin{figure}
    \centering
    \includegraphics[width=0.75\linewidth]{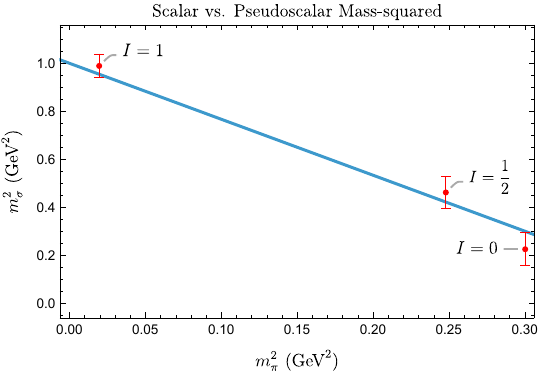}
    \caption{Comparison between the mass-squared scale of different isospin multiplets in the scalar and pseudoscalar spectrum, with isospin labeled. Error bars denote the reported widths of the scalar resonances \cite{ParticleDataGroup:2024cfk}. The blue line corresponds to the $c=7/3$ prediction from ASQCD, with the $m_0$ scale fit as a free parameter. Because the scalar resonances are wide, we cannot consider a precision comparison to the ASQCD prediction. However, the qualitative agreement is surprising and suggests that features of ASQCD, even beyond the vacuum, may be connected to QCD.}
    \label{fig:spectrumfit}
\end{figure}

In particular, we can take our result $c=7/3$ for $N_f=N_c=3$ and compare it to the scalar and pseudoscalar spectrum observed in real-life QCD \cite{ParticleDataGroup:2024cfk}. Perhaps the most immediate implication is the existence of the $0^{+}$ states in QCD such as the $f_0(500)$, which had been controversial for decades before becoming well-established as a real resonance \cite{PELAEZ20161}. It is important to note that the real resonances are broad, attributable to strong coupling. This makes precise comparison to the ASQCD spectrum impossible, but we can ask whether or not there is consistency. In \Cref{fig:spectrumfit} we compare the relative locations of the real scalar and pseudoscalar resonances to the line with slope $-7/3$ and the scale $m_0$ fixed by, for example, the $a_0(980)$ and the $\pi$. We see that the ASQCD result is consistent with the data. One can imagine that in actually taking the $m\to\infty$ limit the ASQCD resonances will broaden into the observed ones.

\section{Conclusion}

In this paper, we presented evidence that the SQCD softly broken by small AMSB $m \ll \Lambda$ (the near-SUSY limit) is continuously connected to the limit of large SUSY breaking $m \gg \Lambda$ (the non-SUSY limit) akin to the BCS--BEC crossover. We showed how non-perturbative condensates of composite operators can be computed for squarks, quarks, gauginos, and gluons. They exhibit a number of non-trivial consistencies between either side of the the theoretically inaccessible region $m \sim \Lambda$, lending plausibility to the notion of continuity. For $N_f < N_c$, their behavior is consistent with the large $N_c$ limit expected in non-SUSY QCD. For $N_f > N_c$, we show how the large $N_f/N_c \sim O(1)$ can modify the large $N_c$ behavior, not possible with previously available methods. We can derive the chiral Lagrangian from first principles for the entire range $N_f < 3N_c$ with a modest assumption of continuity in $N_f$. Surprisingly, we find the low-lying spectrum in the meson superfield to be consistent not only for the $0^-$ pseudo-Nambu--Goldstone bosons (pions) but also for the recently established $0^+$ states. We regard this consistency as a pleasant surprise which strongly supports the hypothesis of a crossover from the near-SUSY to the non-SUSY limit in SQCD. We also confirm that the relationships between the topological susceptibility and the masses of the $\eta'$ and $\pi$, found by Witten and Veneziano in the context of QCD, also hold in ASQCD.

\acknowledgments 
D.\,K. was supported by JSPS KAKENHI Grant Number 24KJ0613. 
The work of H.\,M.\ is supported by the Director, Office of Science, Office of High Energy Physics of the U.S. Department of Energy under the Contract No. DE-AC02-05CH11231, by the NSF grants PHY-2210390 and PHY-2515115, by the JSPS Grant-in-Aid for Scientific Research JP23K03382, MEXT Grant-in-Aid for Transformative Research Areas (A) JP20H05850, JP20A203, Hamamatsu Photonics, K.K, and Tokyo Dome Corportation. In addition, D.\,K and H.\,M.\ are supported by the World Premier International Research Center Initiative (WPI) MEXT, Japan.

\appendix
\section{Calculational Detail of the Kinetic Term}\label{app:kinetic}
In this appendix, we provide the detailed calculation leading to \cref{eq:Qkin}. Before performing the explicit computation, let us first clarify the relation between the kinetic term of the meson field \(U\) and that of \(\xi\). Since the meson field is defined as \(U=\xi^2\), one finds
\begin{align}\label{eq:MesonkineticAppend}
\operatorname{Tr}[\partial_\mu U\partial^\mu U^\dag]
&=\operatorname{Tr}[\partial_\mu (\xi\xi)\partial^\mu (\xi^\dag\xi^\dag)] 
=\operatorname{Tr}[ (\partial_\mu\xi\xi+\xi\partial_\mu\xi) (\partial^\mu \xi^\dag\xi^\dag+\xi^\dag\partial^\mu\xi^\dag)]   \nonumber\\
&=\operatorname{Tr}[ \partial_\mu\xi\xi \partial^\mu \xi^\dag\xi^\dag+\partial_\mu\xi\xi\xi^\dag\partial^\mu\xi^\dag+ \xi\partial_\mu\xi\partial^\mu \xi^\dag\xi^\dag+\xi\partial_\mu\xi\xi^\dag\partial^\mu\xi^\dag)]   \nonumber\\
&=\operatorname{Tr}[ \xi^\dag\partial_\mu\xi\xi \partial^\mu \xi^\dag+\partial_\mu\xi\partial^\mu\xi^\dag+ \partial_\mu\xi\partial^\mu \xi^\dag+\xi\partial_\mu\xi\xi^\dag\partial^\mu\xi^\dag]   \nonumber\\
&=2\operatorname{Tr}[\partial_\mu\xi\partial^\mu\xi^\dag -\xi^2\partial_\mu\xi^\dag \partial^\mu \xi^\dag].
\end{align}
In what follows, we show that two different gauge choices yield the same result.\\ First, let us start from the gauge 
\begin{align}
    Q=\begin{pmatrix}
        v\xi^T\\0
    \end{pmatrix},\ 
    \tilde{Q}=\begin{pmatrix}
        v\xi\\0
    \end{pmatrix}.
\end{align}
Then the kinetic term reads
\begin{align}
    \mathcal{L}_{\text{kin}}
    &=\operatorname{Tr}|D_\mu Q|^2+|D_\mu \tilde{Q}|^2\nonumber\\
    &=v^2\operatorname{Tr}[(\partial_\mu\xi^*-ig\xi^*\rho_\mu^T)(\partial_\mu\xi^T+ig\rho_\mu^T\xi^T)]+v^2\operatorname{Tr}[(\partial_\mu\xi^\dag+ig\xi^\dag\rho_\mu)(\partial_\mu\xi-ig\rho_\mu\xi)]\nonumber\\
    &=v^2\operatorname{Tr}[(\partial_\mu\xi^\dag-ig\rho_\mu\xi^\dag)(\partial_\mu\xi+ig\xi\rho_\mu)]+v^2\operatorname{Tr}[(\partial_\mu\xi^\dag+ig\xi^\dag\rho_\mu)(\partial_\mu\xi-ig\rho_\mu\xi)]\nonumber\\
    &=2v^2\operatorname{Tr}[\partial_\mu\xi^\dag\partial_\mu\xi-ig(\xi^\dag\partial_\mu\xi+\xi\partial_\mu\xi^\dag)\rho_\mu+g^2\rho_\mu^2]\nonumber\\
    &=2v^2\operatorname{Tr}\left[\partial_\mu\xi^\dag\partial_\mu\xi+\left(g\rho_\mu-\frac{i}{2}(\xi^\dag\partial_\mu\xi+\xi\partial_\mu\xi^\dag)\right)^2+\frac{1}{4}(\xi^\dag\partial_\mu\xi+\xi\partial_\mu\xi^\dag)^2\right]\nonumber\\
    &=2v^2\operatorname{Tr}\left[\frac{1}{2}(\partial_\mu\xi^\dag\partial_\mu\xi-\xi^2\partial_\mu\xi^\dag\partial_\mu\xi^\dag)+\left(g\rho_\mu-\frac{i}{2}(\xi^\dag\partial_\mu\xi+\xi\partial_\mu\xi^\dag)\right)^2\right]\nonumber\\
&=2v^2\operatorname{Tr}\alpha_\perp^2+2v^2(g\rho_\mu+\alpha_{\mathbin{\|}})^2,
\end{align}
where
\begin{align}
\alpha_\perp&\equiv \frac{\xi\partial_\mu\xi^\dag-\xi^\dag\partial_\mu\xi}{2i},\ \qquad
\alpha_{\mathbin{\|}}\equiv\frac{\xi^\dag\partial_\mu\xi+\xi\partial_\mu\xi^\dag}{2i}.
\end{align}
We can integrate out the massive ``rho'' mesons $\rho_\mu$ and establish the connection to the meson kinetic term given in~\cref{eq:MesonkineticAppend} as follows.
\begin{align}
    \operatorname{Tr}[\alpha_\perp^2]
    &=-\frac{1}{4}(\xi\partial_\mu\xi^\dag-\xi^\dag\partial_\mu\xi)^2\nonumber\\
    &=-\frac{1}{4}(\xi\partial_\mu\xi^\dag\xi\partial_\mu\xi^\dag-\xi\partial_\mu\xi^\dag\xi^\dag\partial_\mu\xi-\xi^\dag\partial_\mu\xi\xi\partial_\mu\xi^\dag+\xi^\dag\partial_\mu\xi\xi^\dag\partial_\mu\xi)\nonumber\\
    &=\frac{1}{2}\partial_\mu\xi\partial_\mu\xi^\dag-\frac{1}{2}\xi^2\partial_\mu\xi^\dag\partial_\mu\xi^\dag\nonumber\\
    &=\frac{1}{4}\operatorname{Tr}[\partial_\mu U\partial_\mu U^\dag].
\end{align}
Therefore, after decoupling $\rho$, the resulting meson kinetic term becomes
\begin{align}
    \mathcal{L}_{\text{kin}}=\frac{v^2}{2}{\rm Tr} [\partial_\mu U\partial_\mu U^\dag].
\end{align}

Note that integration of ``rho'' mesons is mandatory in some other gauge choices. For instance, with the gauge choice of~\cite{Csaki:2023yas}, parity is not manifest and the ``rho'' mesons actually mix with the pseudoscalar mesons. In this gauge, we take
\begin{align}
    Q=\begin{pmatrix}
        v\\0
    \end{pmatrix},\ 
    \tilde{Q}=\begin{pmatrix}
        vU\\0
    \end{pmatrix}.
\end{align}
Then the kinetic term reads
\begin{align}
\mathcal{L}_{\text{kin}}
    &=\operatorname{Tr}|D_\mu Q|^2+|D_\mu \tilde{Q}|^2\nonumber\\
    &=g^2v^2\operatorname{Tr}[\rho_\mu^2]+v^2\operatorname{Tr}[(\partial_\mu U^\dag+igU^\dag \rho_\mu)(\partial_\mu U-ig\rho_\mu U)]\nonumber\\
    &=g^2v^2\operatorname{Tr}[\rho_\mu^2]+v^2\operatorname{Tr}[\partial_\mu U^\dag\partial_\mu U-2ig\rho_\mu U\partial_\mu U^\dag+g^2\rho_\mu^2 ]\nonumber\\
    &=v^2\operatorname{Tr}\left[\partial_\mu U\partial_\mu U^\dag+2\left(g\rho_\mu-\frac{i}{2}U\partial_\mu U^\dag\right)^2-\frac{1}{2}\partial_\mu U\partial_\mu U^\dag\right]\nonumber\\
    &=\frac{v^2}{2}\operatorname{Tr}[\partial_\mu U\partial_\mu U^\dag]+2v^2\operatorname{Tr}\left[\left(g\rho_\mu-\frac{i}{2}U\partial_\mu U^\dag\right)^2\right].
\end{align}
We thus confirm the agreement between the two expressions for the meson kinetic term, both leading to \cref{eq:Qkin}. We note that a factor of two discrepancy arises if one neglects the contribution of \( \rho_\mu \). We speculate that this is the origin of the factor-of-two difference in the pseudo-scalar mass squared (\( m_{\eta'}^2 \)) compared with the result of~\cite{Csaki:2023yas}. By carefully integrating out \( \rho_\mu \), as shown here, one obtains a factor-of-two enhancement in \( m_{\eta'}^2 \) discussed in \eqref{eq:metap2} and the foonote associated with it.


\section{Trace Calculation Up To Local Counter Term}\label{sec:WZWtracecalc}
In this section, we show the equality of the WZW term and the Jacobian of fermion determinant up to a local counter term. Recall that anomalies are ambiguous without imposing remaining symmetries, and different forms of anomalies are related by adding local counter terms. See, {\it e.g.}\/, \cite{Fujikawa:2003az} for a dedicated discussion about local counter terms for anomalies. In our case, the WZW term is derived from integrating out massive fermions in the spectrum, yet it originally does not have the familiar form. Here we demonstrate that the difference can indeed be accounted for with a local counter term. 

We use $U=\xi^2$ as a matrix. Then
\begin{align}
    U^\dag dU &= \xi^\dag (\xi^\dag d\xi+d\xi \xi^\dag)\xi,\nonumber\\
    \operatorname{Tr}[(U^\dag dU)^5]
    &=\operatorname{Tr}[(\xi^\dag d\xi+d\xi \xi^\dag)^5] 
    =\operatorname{Tr}[(\xi^\dag d\xi-\xi d\xi^\dag)^5]. 
\end{align}
Here, we set $A=\xi^\dag d\xi$, $B=-\xi d\xi^\dag$. We can see the following relations
\begin{align}
    dA&= -A^2,\ 
    dB=B^2.
\end{align}
The WZW term is in the following form
\begin{align}
    \operatorname{Tr}[(A+B)^5]
    =\operatorname{Tr} [ A^5+ 5A^4B+ 5A^3B^2+ 5A^2BAB+5A^2B^3+ 5ABAB^2+ 5AB^4+ B^5 ].
\end{align}
By replacing higher order term by derivative using above relations, we can obtain
\begin{align}
\operatorname{Tr} [(A+B)^5]
&=  \operatorname{Tr}  \left [A^5+B^5+ 5d(AdAB)- \frac{5}{2} d (AB)^2   -5d(AdBB) \right] \nonumber \\
&=  \operatorname{Tr}  \left [A^5+B^5\right]
+ d\operatorname{Tr} \left[5(AdAB)- \frac{5}{2} (AB)^2   -5(AdBB) \right]. 
\end{align}
Since the WZW term is written as an integral over a five-dimensional manifold whose boundary is identified with our four-dimensional spacetime, the exact form in the second term above is a local expression in four dimensions. 
Therefore, we can see the equality of Jacobian and WZW term up to local counter terms.

\section{Origin of Factor $N_c$ in the WZW Term}\label{sec:originNc}

In this section, we identify the origin of the overall factor $N_c$ in the Wess--Zumino--Witten (WZW) term in the chiral Lagrangian in our derivation of the chiral Lagrangian in \cref{sec:chiralLagrangianADS}. It arises from two independent contributions from the Jacobians in the path integral when the fermion mass matrix is diagonalized. We specifically study the ADS case.

Recall the form of the VEVs in \cref{eq:ADSvevs}. The fermion components split into the upper $N_f\times N_f$ block and the lower $(N_c-N_f)\times N_f$ block,
\begin{align}
    Q = \left( \begin{array}{c} v \xi^T + \theta \psi \\
    \theta \chi \end{array} \right), \qquad
    \tilde{Q} = \left( \begin{array}{c} v \xi + \theta \tilde{\psi} \\ \theta\tilde{\chi} \end{array} \right) .
\end{align}
(Here, we dropped the $F$-components we are not interested in.)
Their transformation laws under $g \in G_{\rm global}$ are
\begin{align}
    Q \rightarrow \left( \begin{array}{c} v h \xi^T g_L^T + \theta h \psi g_L^T \\
    \theta \chi g_L^T \end{array} \right), \qquad
    \tilde{Q} \rightarrow \left( \begin{array}{c} v h^* \xi g_R^T + \theta h^* \tilde{\psi} g_R^T \\ \theta\tilde{\chi} g_R^T \end{array} \right) .
\end{align}
On the other hand, the gaugino fields also split into blocks
\begin{align}
    \lambda = \left( \begin{array}{cc}
        \lambda_{11} & \lambda_{12} \\
        \lambda_{21} & \lambda_{22}
    \end{array} \right)
    \rightarrow \left( \begin{array}{cc}
        h \lambda_{11} h^\dagger & h \lambda_{12} \\
        \lambda_{21} h^\dagger & \lambda_{22}
    \end{array} \right).
\end{align}

The scalar-fermion-gaugino interaction induces the mass terms
\begin{align}
    \int d^4 \theta {\rm Tr} Q^\dagger e^V Q
    \supset v {\rm Tr} (\xi^* \lambda_{11} \psi)
    + v {\rm Tr} (\xi^* \lambda_{12} \chi) + {\rm h.c.},
\end{align}
and
\begin{align}
    \int d^4 \theta {\rm Tr} \tilde{Q}^\dagger e^{-V^T} \tilde{Q}
    \supset v {\rm Tr} \xi^\dagger \lambda_{11}^T \tilde{\psi}
    + v {\rm Tr} \xi^\dagger \lambda_{21}^T \tilde{\chi}
    + {\rm h.c.},
\end{align}
which are indeed invariant under $G_{\rm global}$. 

On the other hand, the ADS superpotential depends on
\begin{align}
    \tilde{Q}^T Q
    &\supset \left( \begin{array}{cc}
        v^2 \xi^T \xi^T + v \theta \xi^T \psi + v \theta \tilde{\psi}^T \xi^T + \theta^2 \tilde{\psi}^T \psi + \theta^2 \tilde{\chi}^T \chi
    \end{array} \right) \nonumber \\
    &= \xi^T \left( \begin{array}{cc}
        v^2 + v \theta \psi \xi^* + v \theta \xi^* \tilde{\psi}^T + \theta^2 \xi^* \tilde{\psi}^T \psi \xi + \theta^2 \tilde{\chi}^T \chi
    \end{array} \right) \xi^T .
\end{align}
The mass matrix can only depend on the combinations
\begin{align}
    {\rm Tr} (\psi \xi^* \xi^* \tilde{\psi}^T), \quad
    {\rm Tr} \tilde{\chi}^T \chi.
\end{align}
In order to diagonalize the mass matrix, we need to perform the reparametrization
\begin{align}
    \psi \rightarrow \psi \xi^T, \qquad
    \tilde{\psi} \rightarrow \tilde{\psi} \xi,
\end{align}
yielding a factor of $N_f$ for the Jacobian, while anothe reparameterization
\begin{align}
    \chi \rightarrow \chi \xi^T, \qquad
    \tilde{\chi} \rightarrow \tilde{\chi} \xi,
\end{align}
yielding a factor of $N_c - N_f$ for the Jacobian. Combining them together, there is an overall factor of $N_c$.

\bibliographystyle{JHEP}
\bibliography{ref}
\end{document}